\documentstyle{l-aa}
\begin{document}

   \thesaurus{ 11.19.3; % Galaxies: starburst
               13.09.1; % Infrared: galaxies
               11.19.5; % Galaxies: stellar content
               11.05.2; % Galaxies: evolution
                }

   \title{The history of star formation of starburst galaxies}

   \author{R. Coziol
           \inst{1}
                }

   \offprints{R. Coziol, coziol@das.inpe.br}

   \institute{Divis\~ao de Astrof\'\i sica,
              Instituto Nacional de Pesquisas Espaciais,
              CP~515, 12201--970
              S\~ao Jos\'e dos Campos, Brazil
                }

   \date{Received   ; accepted   }

   \maketitle

\begin{abstract}
In this paper, we use different luminosity ratios to trace the
histories of star formation of two different kinds of starburst
galaxies: HII galaxies and starburst nucleus galaxies (SBNGs). The
mean star formation rates (SFRs) for these galaxies is comparable,
and compatible with a near constant star formation over a few Gyr
period. We have also found an interesting difference between the
burst stellar population of the SBNGs and the HII galaxies: SBNGs
have an excess of intermediate-mass stars as compared to HII
galaxies. We interpreted this difference as a sign that SBNGs have
experienced a higher frequency of bursts of star formation than
the HII galaxies. This interpretation is qualitatively consistent
with the Stochastic Self Propagation of Star Formation (SSPSF) theory,
which suggests that the starburst phenomenon depends on internal
processes, regulated by the evolution of massive stars.

\end{abstract}
%
%________________________________________________________________

\section{Introduction}

Peculiar galaxies with unusual level of star formation are commonly
known as starburst galaxies.  The definition of a starburst
galaxy is however rather vague, and mostly model dependent.
The ``standard'' model proposes that the typical time scale of the
starburst phenomenon is of the order of $10^7$ to $10^8$ years.
Short time scales are assumed because young massive stars, which
dominate the optical spectrum of a starburst galaxy, cannot live
very long, and high star formation rates will theoretically exhaust
any available gas reservoir very rapidly. Short time scales
correspond well also to typical time scales of interacting
galaxies, which is the favored scenario to explain the origin of
the starburst.

On observing ground, the standard model is far from being conclusive.
For example, a short time scale for the bursts implies that we
should see a lot of post-burst galaxies, that is galaxies with
E+A spectral type. But, those are rarely found (Norman 1991). The
standard model do not explain also the variety of starburst
galaxies observed  (Salzer et al. 1989; Campos-Aguilar \& Moles 1991). In
particular, it does not explain what could be the difference between
the two main groups of starburst galaxies: the HII galaxies
(Terlevich et al. 1991) and the Starburst Nucleus Galaxies (SBNGs;
Balzano 1983; Coziol et al. 1994). HII galaxies are usually located
in small, metal poor galaxies, with low dust content (Pe\~na et al.
1991; Coziol et al. 1994; Masegosa et al.  1994). By comparison, SBNGs
are located in more chemically evolved and massive galaxies, with a
large population of old or evolved stars and a huge quantity of dust
(Mirabel \& Duc 1993; Coziol et al. 1994; Coziol et al. 1995a). In
general, HII galaxies are either irregular galaxies or blue compact
dwarfs (BCDs), while SBNGs are more numerous among early-type
galaxies (Sb and earlier; Coziol et al. 1995a). The standard model of
starburst is furthermore inconclusive  because the interaction
hypothesis is not always confirmed.  In fact, for a great number of
starburst galaxies direct signs of interaction are either missing or
the galaxies are simply isolated (Taylor et al. 1993; Telles \&
Terlevich 1995; Barth et al. 1995; Coziol et al. 1995b).

\begin{figure*}
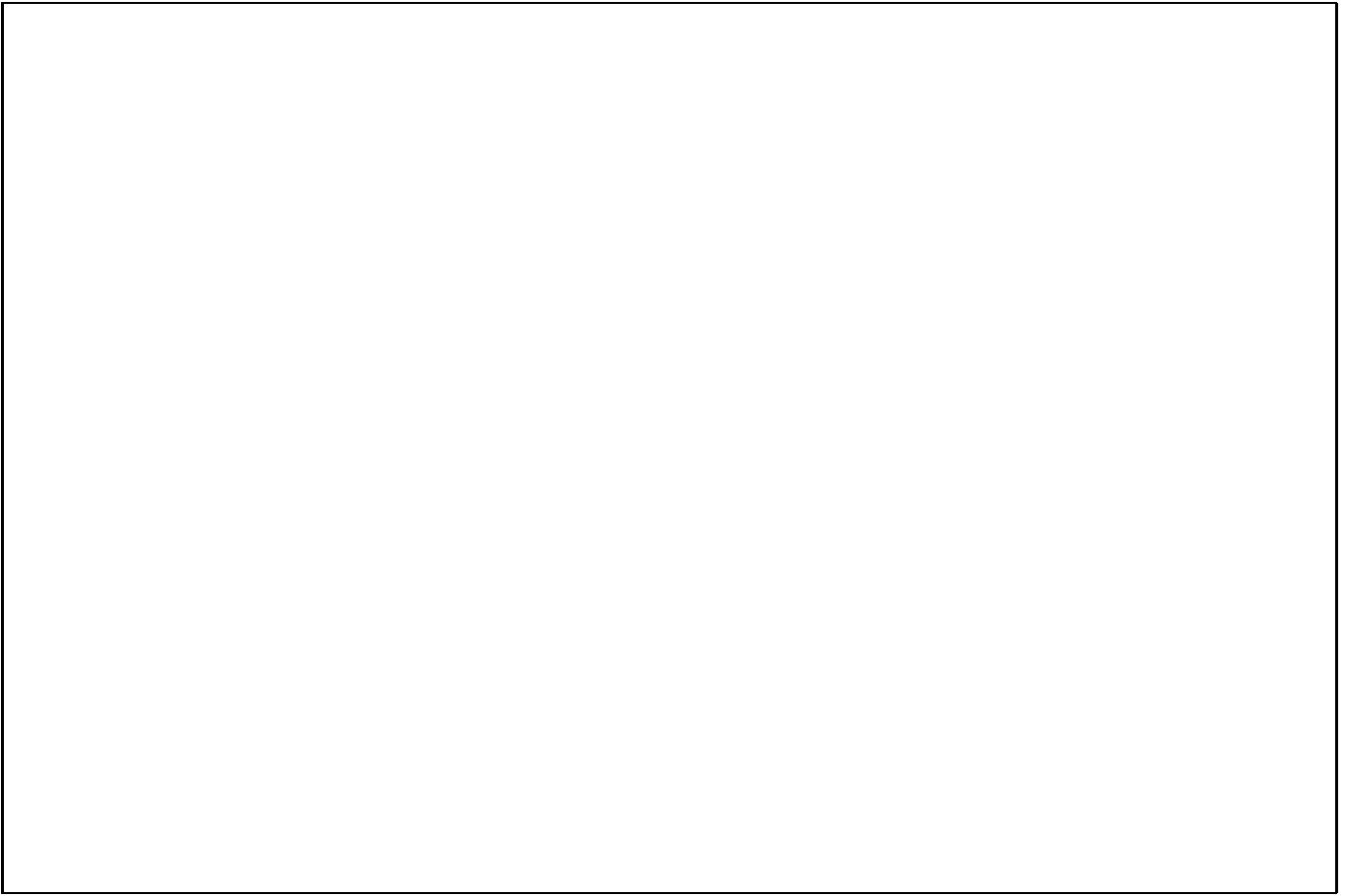

\picplace{12cm}
\caption{ Diagnostic diagram of the different kind of starburst
galaxies: the high--excitation HII galaxies and the low--excitation SBNGs.
The frontier between the two starburst types is placed at Log([OIII]/H$\beta$)$
\geq 0.4$.
Compared to HII regions models, the different starbursts trace a quasi
continuous
sequence in metallicities.}
\label{Fig.1}
\end{figure*}

The goal of this paper is to gain a wider view of the starburst
phenomenon based on their histories of star formation. To do that,
we compare different luminosity ratios which are taken as tracers of
star formation rates (SFRs) over different periods of time
(Kennicutt 1983; Gallagher et al. 1984; Thronson \& Telesco 1986).
By comparing large and different samples of galaxies, we can both
develop a coherent view of the evolution of the star formation in
galaxies and arrive to a better definition of the starburst
phenomenon.

\section{The different types of starburst galaxies}

In Fig. 1, we compare different samples of galaxies by the mean of
the diagnostic diagram of [NII]$\lambda$6584/H$\alpha$ against
[OIII]$\lambda$5007/H$\beta$ (Baldwin et al. 1981; Veilleux \&
Osterbrock 1987). In this figure, the sample of IRAS galaxies comes
from Allen et al. (1991), the sample of HII galaxies from Pe\~na et al.
(1991), and the SBNGs from the Montreal Blue Galaxy (MBG) survey
(Coziol et al. 1993; Coziol et al. 1994). In this diagram, the
continuous curved marked AGNs is the empirical separation between
galaxies where the gas is assumed to be ionized by stars, and
galaxies where the main ionizing source is a power law (Veilleux \&
Osterbrock 1987). The curve marked HII regions is the model of disk
HII regions developed by Evans \& Dopita (1985).

The main feature of this diagram is that the two types of starburst galaxies
can be distinguished based on their level of excitation: the HII
galaxies show high--excitation spectra and the SBNGs show
low--excitation spectra. The frontier between the two groups is
situated at Log([OIII]/H$\beta$)$ \geq 0.4$. This limit is almost
the same as the one, proposed by Shuder \& Osterbrock (1981), to
separate AGNs from LINERs (that is [OIII]/H$\beta \geq 3$).
In Fig.~1, the high-excitation starburst galaxies are easily
distinguished from the AGNs. But this is not the case for the
low-excitation galaxies, and there appears to be a good number of
galaxies with intermediate spectroscopic characteristics between a
starburst and a LINER.  The ambiguous nature of these galaxies comes
from the fact that in SBNGs the mean ratio of Log([NII]/H$\alpha$) is
0.2 dex higher than the mean ratio predicted by normal HII region
models (Coziol et al. 1995b). At the moment, it is still unclear if
this phenomenon is solely the consequences of dust in metal-rich
HII regions (Shields \& Kennicutt 1995) or if it reflects a real
composite nature of these galaxies (Kennicutt et al. 1989).

\begin{table*}
\caption{Two component models for the far far--infrared radiation}
\begin{tabular}{lccccc} \hline\noalign{\smallskip}
&HII galaxies&SBNGs&late-type&early-type\\
\noalign{\smallskip}\hline\noalign{\smallskip}
Number&143&94&79&124\\
$<f_{60}/f_{100}>$&$0.57 \pm 0.21$&$0.52 \pm 0.18$&$0.38 \pm 0.11$&$0.29 \pm
0.40$\\
\noalign{\smallskip}\hline
\end{tabular}
\end{table*}

In Fig. 1, the relative scarcity of starburst galaxies in the
transition region $-0.9 <$ Log([NII/H$\alpha$)$ < -0.6$ can be
explained in part by the observational biases of the different
surveys. Objective-prism surveys are biased against low-excitation
galaxies (Masegosa et al.  1994), while color surveys, like the
MBG survey, are biased against high-excitation galaxies (Coziol et
al. 1994). Similarly, very few IRAS galaxies are detected in the
transition region because HII galaxies have a lower dust content
than SBNGs. Therefore, there may be a continuous sequence of
excitation of starburst galaxies.
Theoretically, such continuous sequence of excitation could be
explained by HII regions photoionization models (McCall et al.
1985; Evans \& Dopita 1985; Campbell 1988; Pe\~na et al. 1991;
McGaugh 1991). Following these authors, the most important
parameter is the abundance of elements: metal poor galaxies produce
higher excitation than metal rich galaxies. On this matter, it is not
clear that the small number of galaxies with
Log([NII]/H$\alpha$)$ < -1.8$ can be explained by a detection bias effect
(McGaugh
1991; Masegosa et al. 1994). In fact, this phenomenon more probably
reflects the rarity of galaxies with very low metallicities (Campbell
1988; Pe\~na et al. 1991).  In this sense, galaxies resembling IIZW40
should be consider rather special.

The fact that HII galaxies are less chemically evolved than SBNGs
are usually interpreted as a consequence of different types of
phenomena. Among several hypotheses, HII galaxies could be either
young galaxies (Terlevich et al. 1991) or galaxies with retarded
star formation (Searle \& Sargent 1972; Taylor et al. 1993),
whereas SBNGs are considered as old galaxies, which were
rejuvenated by a recent infall of matter (Huchra 1977). Following
these hypotheses, it is not obvious why we should see a continuous
sequence of metallicity of the different starburst galaxies.
Different phenomena should also imply different histories of star
formation.

\section{The different tracers of star formation rates}

We can follow the history of star formation of galaxies by tracing their SFRs
on two  different time scales (Kennicutt 1983, Gallagher et al.
1984, Thronson \& Telesco 1986).
The first tracer of star formation is the H$\alpha$
luminosity (L$_{H\alpha}$). L$_{H\alpha}$ is related to massive OB
stars and therefore reflects the recent SFR, with time scale of
$10^6$--$10^7$ yrs. The second tracer of star formation is the B luminosity
(L$_B$).
L$_B$ traces the SFR on time scale of $4 \times 10^8$ to $6 \times
10^9$ yrs, when stars of masses between 1 and 3 M$_\odot$ dominate
the main sequence (Larson \& Tinsley 1978; Gallagher et al. 1984).

\begin{figure}
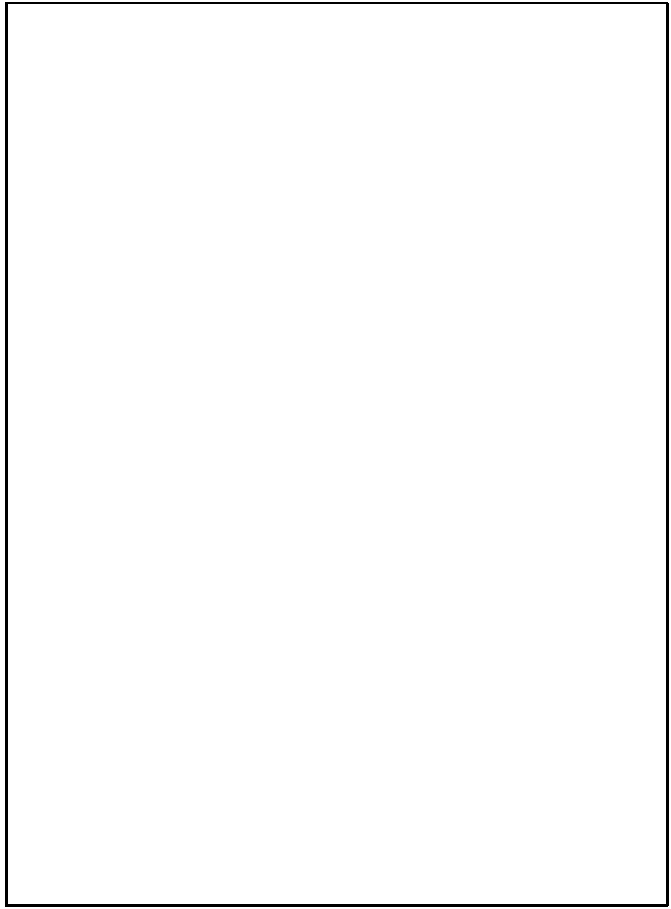

\picplace{12cm}
\caption{Comparison of the $f_{60}/f_{100}$ IRAS ratios for the two
types of starburst galaxies with those found in normal spirals.
The high ratios for the starbursts suggest that in those
galaxies the far--infrared luminosity is related to young massive stars.}
\label{Fig.2}
\end{figure}

There is two problems with L$_{H\alpha}$. The first one is that
L$_{H\alpha}$ is sensitive to dust extinction, which effects
are still poorly understood in external galaxies. The second problem
is that there still are very few galaxies with H$\alpha$
fluxes measured. For these reasons, but also because of the
of the IRAS survey, it was proposed to use the far--infrared
luminosity (L$_{IR}$) as a tracer of recent SFR (Thronson \&
Telesco 1986). The idea is that active star forming regions are
associated to a large quantity of dust which is predominantly
heated by the massive young stars. Unfortunately, there is also a
controversy on this matter and some authors have warned against the
generalized use of L$_{IR}$ as a tracer of recent SFR (Sauvage \&
Thuan 1992). One of the problems is that we do not know what is the
contribution of older population stars in heating the dust.
However, we think it is well demonstrated that in active star
forming galaxies the contribution of young massive stars dominates
in the far--infrared (Helou 1986; Sekiguchi 1987).

\begin{table*}
\caption{Statistical characteristics}
\begin{tabular}{lccccc}
\hline\noalign{\smallskip}
&SBNGs&HII galaxies&NGC late-type&HII nucleus galaxies&NGC early-type\\
\noalign{\smallskip}\hline\noalign{\smallskip}
Number of objects           & 143            & 61             & 79
& 79             & 121\\
Log$<$L$_{IR}>$   & $9.87 \pm 0.70$& $8.88 \pm 0.84$& $9.50 \pm 0.52$& $9.41
\pm 1.22$& $8.92 \pm 0.62$\\
Log$<$ L$_B>$     & $9.70 \pm 0.54$& $8.52 \pm 0.61$& $9.65 \pm 0.39$& $9.91
\pm 0.35$& $9.70 \pm 0.45$\\
Number of objects & 90             & 50             & 79             & 79
      & 0\\
Log$<$L$_{H\alpha}>$ & $6.90 \pm 0.67$& $7.04 \pm 0.74$& $7.26 \pm 0.43$& $5.61
\pm 0.67$& --\\
Correlation coefficients:&&&&&\\
L$_{IR}$ vs. L$_B$         & 0.84& 0.76& 0.79& 0.39& 0.54\\
L$_{H\alpha}$ vs. L$_B$    & 0.72& 0.78& 0.70& 0.29& --\\
L$_{H\alpha}$ vs. L$_{IR}$ & 0.87& 0.93& 0.77& 0.29& --\\
Slope of the linear regressions:&&&&&\\
L$_{IR}$(L$_B$)   & $1.10 \pm 0.06$& $1.05 \pm 0.11$&  $1.06 \pm 0.10$&  $1.36
\pm 0.36$& $0.74 \pm 0.11$\\
L$_{H\alpha}$(L$_B$)   & $0.92 \pm 0.10$& $0.95 \pm 0.11$&  $0.58 \pm 0.07$&
$0.54 \pm 0.19$&  --\\
L$_{H\alpha}$(L$_{IR}$)& $0.82 \pm 0.04$& $0.82 \pm 0.07$&  $0.59 \pm 0.06$&
$0.33 \pm 0.12$&  --\\
Luminosity ratios:&&&&&\\
Number of objects             &  88 &  38 & 79  & 79  &121\\
Log$<$ L$_{IR}/$L$_B>$        &  0.2&  0.1& -0.2& -0.4& -0.8\\
Log$<$ L$_{H\alpha}/$L$_B>$   & -2.8& -2.2& -2.4& -4.4&  --\\
Log$<$ L$_{IR}/$L$_{H\alpha}>$&  3.0&  2.2&  2.2&  4.0&  --\\
\noalign{\smallskip}\hline
\end{tabular}
\end{table*}

As a further test, we can compare the two IRAS flux at 60 $\mu$m and
100 $\mu$m ($f_{60}/f_{100}$). Following Mazzarella et al. (1991)
this ratio is a good indicator of the nature of the dust heating
sources in galaxies. It provides a direct indication of the
dominance of the warm component of the interstellar radiation field
produce by OB stars as compared to the cooler cirruslike component
due to old stars. In Fig. 2, we compare the distribution of the
$f_{60}/f_{100}$ ratios for the SBNGs and the HII galaxies with
those of two samples of NGC normal field galaxies. The early-type
galaxies are from Hogg et al. (1993).  This last sample was chosen
because it represents a group of galaxies with low SFRs (Hogg et
al. 1993). The sample of late-type galaxies are from Kennicutt
(1983). This sample is composed mostly of late-type spiral galaxies.
It represents normal star formation in the disk of galaxies. Mean values of the
$f_{60}/f_{100}$ ratios for the different types of galaxies are
reported in Table 1. In Fig. 2 it can be seen that the
HII galaxies and the SBNGs have similar distributions. Both
starburst galaxies have comparable mean $f_{60}/f_{100}$ ratios
(see Table 1). Furthermore, the mean ratios for both types of
starburst galaxies are clearly higher than the mean ratio for the
late-type spirals. Therefore in starburst galaxies the far--infrared
luminosity is mostly related to the young massive stars.

{}From Fig. 2, we can see that the
distribution of the $f_{60}/f_{100}$ for the NGC field galaxies
follow the well known variation of star formation along the Hubble
types: the SFRs in late-type galaxies are higher than in early-type
galaxies. In Fig. 2, the increases of the SFRs along the Hubble
sequence is also correlated in the far--infrared to an increases of
the mean ratio of $f_{60}/f_{100}$.  Note that from the original
sample of 279 early-type galaxies considered above, only 44\% were
detected by IRAS. By comparison, 79\% of the sample of the
late-type galaxies were detected. This phenomenon reflects the fact
that the most active star forming galaxies are also the dust
richest. If we consider those galaxies as closed systems, the
amount of dust a galaxy contain should be related in some way to
its history of star formation. Whence, the low dust content of
early-type galaxies is also correlated to their present low star
formation rates.

\subsection{Relations between the different luminosities}

In order to convince ourselves of the validity of our different
tracers of star formation, we now compare L$_{H\alpha}$ with
L$_{IR}$ (Fig. 3) and L$_B$ (Fig. 4). Table 2 gives a summary of the
different relations found by comparing the luminosities.  Five
different types of galaxies are considered. The HII galaxies are
composed of galaxies from the UM survey (Salzer et al.  1989) and
the Calan-Tololo survey (Pe\~na et al. 1991). The SBNGs are taken
from the Balzano's sample (1983) and from the MBG survey (Coziol et
al. 1993; Coziol et al. 1994). The sample of NGC late-type spirals
are from Kennicutt (1983). The fourth sample is the HII nucleus
galaxies. Those are defined as galaxies with HII regions in their
nucleus (Kennicutt et al. 1989). Together with LINERs galaxies, the
HII nucleus galaxies are the most frequent types of galaxies found
in spectroscopic surveys (French 1980; Heckman 1980; Stauffer
1983). In terms of star formation, the HII nucleus galaxies could
represent the low-luminosity end of SBNGs.  The last sample is
composed of the early-type galaxies from Hogg et al. (1993).

\begin{figure}
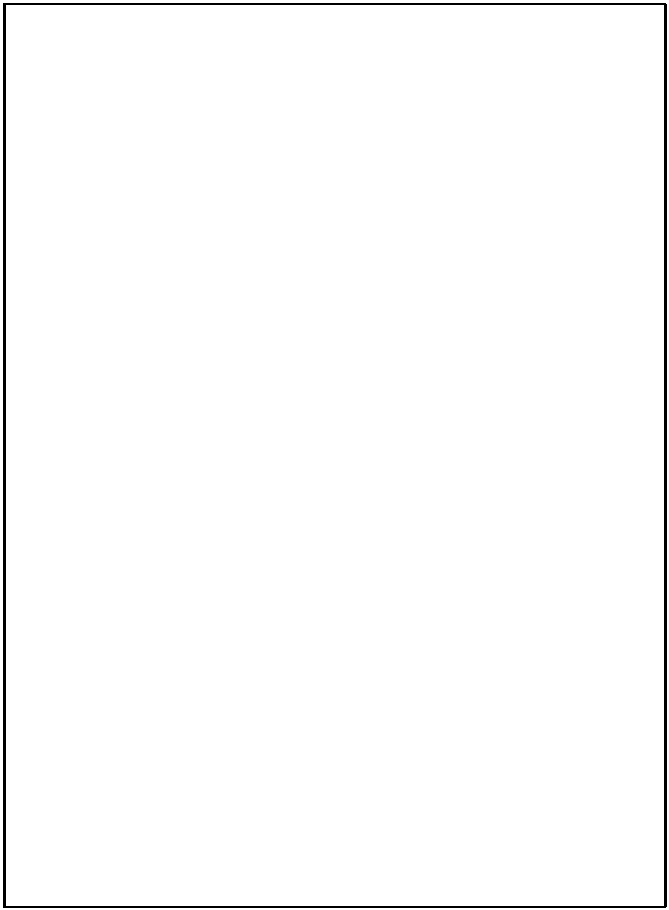

\picplace{12cm}
\caption{Relation between L$_{H\alpha}$ and L$_{IR}$. In the active
star--forming galaxies
the far--infrared luminosity is correlated to the H$\alpha$ luminosity. For the
SBNGs,
the slope of the linear relation is the same as for the HII galaxies, but the
intercept is different.}
\label{Fig.3}
\end{figure}

\begin{figure}
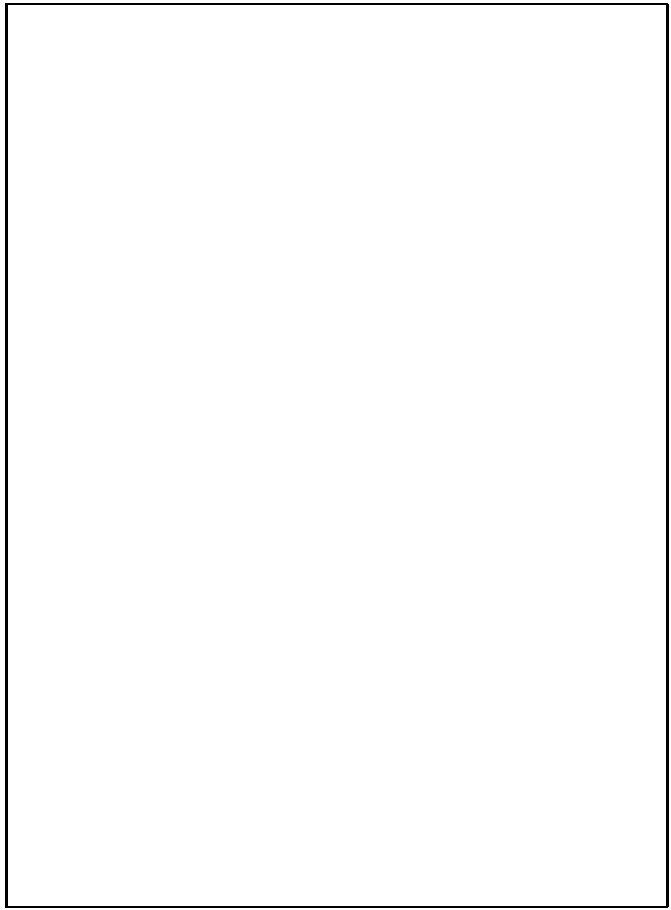

\picplace{12cm}
\caption{For all the active star--forming galaxies, we find a poorer
correlation between
L$_{H\alpha}$ and L$_B$.  For the SBNGs the correlation is slightly better
than for the HII galaxies or than for the late--type spirals. For the two types
of starbursts
the slopes of the linear relations are the same, but not the intercepts. }
\label{Fig.4}
\end{figure}

The different luminosities were evaluated using heliocentric
velocity ($V_h$) corrected for the motion of the sun according to
$V_o = V_h + 250$ sin($l$) cos($b$), where $l$ and $b$ are the
galactic longitude and latitude of the objects. In this paper, we
use $H_o = 75$ km s$^{-1}$ Mpc$^{-1}$. The $H\alpha$ fluxes were
taken from the original articles, and are uncorrected for intrinsic
extinction.  Unfortunately, no H$\alpha$ fluxes are available for the
early-type galaxies.
The coordinates, the redshifts, the B magnitudes and the
IRAS fluxes were all taken from the NASA/IPAC Extragalactic Database
(NED). The B luminosity were corrected for Galactic extinction, as found in
NED. The
far--infrared luminosity is given by Log(L$_{IR}$) = Log($F_{IR}$) + 2
Log[z(z+1)] + 57.28, where z is the redshift and $F_{IR}$ = $1.26
\times 10^{-11}(2.58f_{60}+f_{100})$ erg cm$^{-2}$ s$^{-1}$
(Londsdale et al. 1985).

Fig. 3 suggests that in the HII galaxies, the SBNGs and the
late-type spirals, L$_{H\alpha}$ is linearly correlated to
L$_{IR}$. This relation is lost for the HII nucleus galaxies.  On
the other hand, Fig. 4 suggests a poorer correlation between
L$_{H\alpha}$ and L$_B$.  Considering the correlation coefficients
in Table 2, we see that the correlation between L$_{H\alpha}$ and
L$_{IR}$ and L$_{H\alpha}$ and L$_B$ follow a decreasing sequence:
HII galaxies, SBNGs, disk HII regions, HII nucleus galaxies, with
the exception that the far--infrared luminosity is better correlated to
the B luminosity in SBNGs. Strictly speaking, no correlations are
found between the different luminosities in the HII nucleus
galaxies. This behavior is consistent with our interpretation of
the luminosities in terms of different tracers of star formation:
the far--infrared luminosity, like the H$\alpha$ luminosity, is
mostly related to the young stellar population, while the B
luminosity is mostly related to intermediate-mass stars.

\subsection{Size effect and the role of dust in
luminosity-luminosity diagrams}

\begin{figure*}
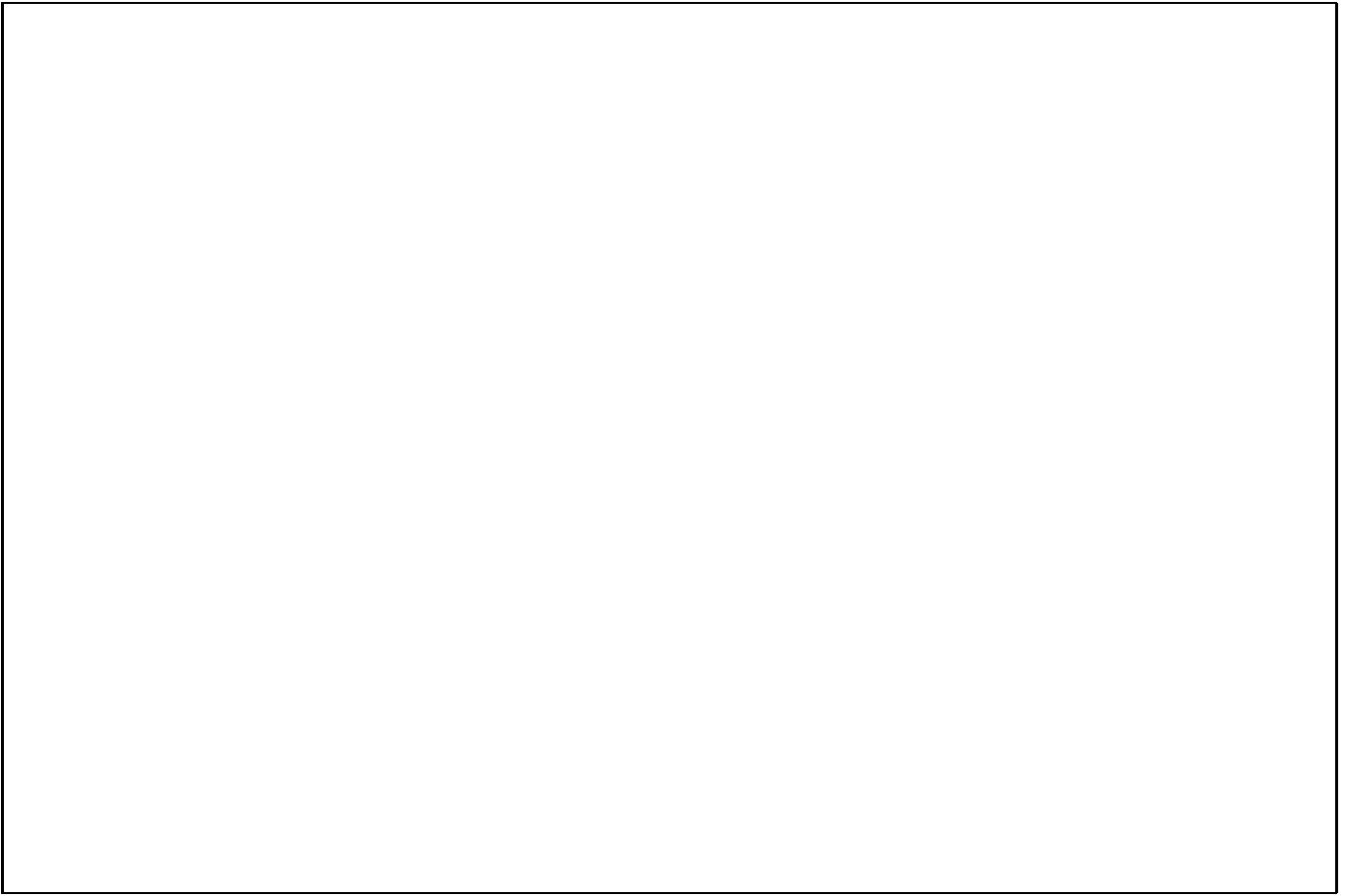

\picplace{12cm}
\caption{Normalized luminosity--luminosity diagram. Taking into account the
size effect, active
star--forming galaxies show a correlation of the far--infrared luminosity with
the H$\alpha$ luminosity.
The SBNGs have lower ratios of ionized gas per unit of stellar mass. This
cannot be explained by
higher dust extinction, because the ratios of dust per unit of stellar mass are
the same for all the active star--forming galaxies.}
\label{Fig.5}
\end{figure*}

In Fig. 3, the fact that L$_{H\alpha}$ increases with L$_B$
suggests that the different correlations found between the
luminosities are affected by a ``size effect³³. To take into account
this effect, we have compared the ratios L$_{H\alpha}$/L$_B$ with
L$_{IR}$/L$_B$, where L$_B$ is taken here as an estimator of the
stellar mass of the galaxies. In Fig. 5, we can still see a linear
increase of the far--infrared emission with the emission in
H$\alpha$. These relations suggest that the amount of far--infrared
emission per unit of stellar mass increases with the amount of
ionized gas per unit of stellar mass. This behavior is a common
characteristic of all the active star forming galaxies.

By comparing the luminosity ratios of the two different types of
starburst galaxies, one interesting difference can be seen in all
the figures: the SBNGs do not follow exactly the same relations as
the HII galaxies. For the two types of starburst galaxies, the
relations have the same slopes, but not the same intercept. In
Fig.  5, this difference corresponds to a lower amount of ionized
gas per unit of stellar mass in the SBNGs than in the HII
galaxies.  The simplest explanation for this phenomenon is to
suppose an higher dust extinction in SBNGs. However, an higher
extinction in SBNGs would also contribute to destroy any relation
between the blue luminosity and the far--infrared luminosity, but in
fact the correlation between these luminosities is better in SBNGs
(see Table 2). Actually, if extinction was an important factor, one
would have expected to see in SBNGs the far--infrared luminosity
increasing faster than the H$\alpha$ luminosity. The hypothesis of
higher dust extinction in SBNGs seems also incoherent with the fact
that both types of galaxies have the same mean far--infrared
luminosities per unit of stellar mass, and the same mean dust
temperature as determined by the comparable ratios of
$f_{60}/f_{100}$.

In their study of starburst galaxies, Calzetti et al. (1995;
hereafter CBKSH) have found a linear relation between the
extinction in the optical, as extrapolated from the UV band, and
the ratio L$_{IR}$/L$_B$. Following their interpretation, an
increase of dust in a galaxy corresponds to a lower B luminosity and
to a larger far--infrared emission. Applying their interpretation to Fig.
5, the amount of dust extinction in the optical increases from 0 at
Log(L$_{IR}$/L$_B$) = -1.0 to 3 magnitudes at Log(L$_{IR}$/L$_B$) =
+1.0. As one can see in Fig. 5, this interpretation cannot explained
the differences between the two types of starbursts, because both
span the same range of L$_{IR}$/L$_B$. Furthermore, the relation
shown in Fig. 5, goes opposite to what we should expect from
extinction effect: the slopes should be negatives.  In Fig.  5, the
behavior of L$_{IR}$/L$_B$ is better interpreted in terms of
different SFRs: the SFR increases with L$_{IR}$/L$_B$. The relation
found by CBKSH between optical extinction and L$_{IR}$/L$_B$ simply
reflects the fact that the most active star forming galaxies are
also the dustiest, as we have observed before. Therefore, the
difference between SBNGs and HII galaxies must corresponds to
another phenomenon.  Fig. 5 furthermore suggests that this
difference should vanish in the more active star forming galaxies.

{}From Table 2, we see that the correlation between L$_{H\alpha}$ and
L$_{IR}$ is poorer in the late-type spirals.  However, those
galaxies have comparable SFRs as the SBNGs or the HII galaxies. The
mean temperature of the dust is also lower in the late-type spirals
than in the two type of starburst galaxies. One possible
explanation is because the star formation in the late-type galaxies
happens only in their spiral arms. Consequently, the active star
forming regions are more dispersed and the contribution of older
stellar population is more obvious in the B band and the
far--infrared. This scenario is furthermore supported by the
different slopes of the linear regressions, where we can see that
both the B luminosity and the far--infrared luminosity increase
faster than the H$\alpha$ luminosity. In the case of the HII
nucleus galaxies, the fact that we do not see any  correlations
between the luminosities, is clearly related to their low mean
H$\alpha$ luminosity. This could only mean that the SFRs on these
galaxies is much lower than in the late-type spirals or that in the
starburst galaxies. This interpretation is furthermore coherent
with their mean ratio $f_{60}/f_{100}$, which is comparable to
galaxies where the dust is mostly heated by old stars.

\section{The history of star formation of starburst galaxies}

\begin{figure*}
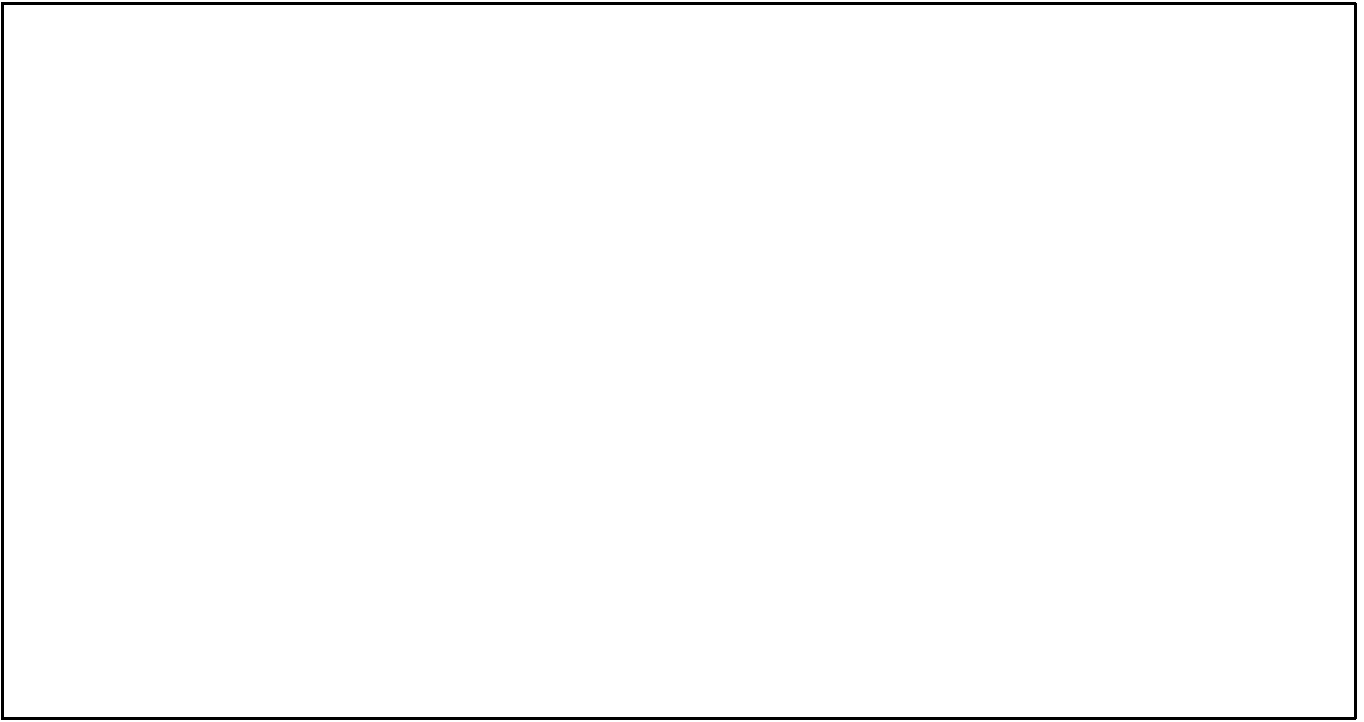

\picplace{9.5cm}
\caption{Calibration of the L$_{IR}$ vs. L$_B$ diagram in terms of SFRs.
The diagonals correspond to different ratios of
L$_{IR}$/L$_B$. Most of the late-type spirals have ratios $1/3 <$
L$_{IR}$/L$_B$ $< 3$. The
galaxies in this region have a near constant star formation rate over a 3 Gyr
period.}
\label{Fig.6}
\end{figure*}
\begin{figure*}
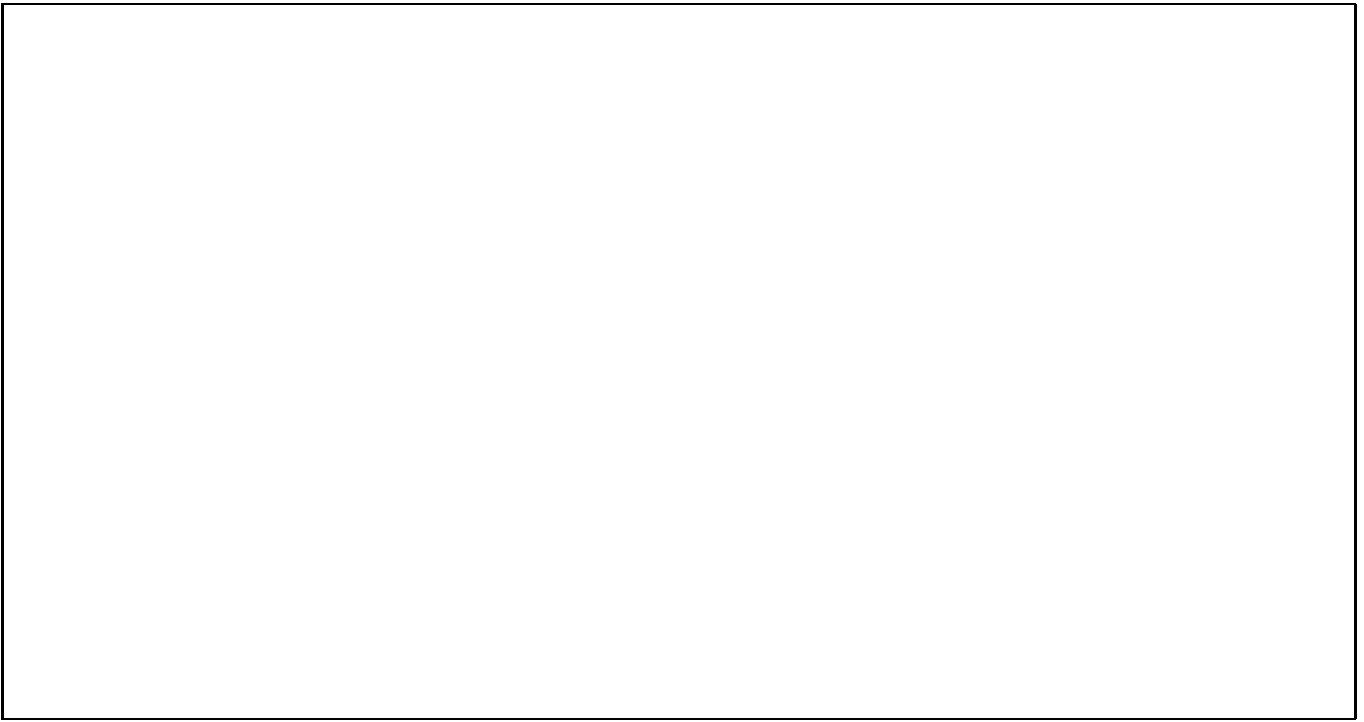

  \picplace{9.5cm}
  \caption{The history of star formation of starburst galaxies. Very few
starburst galaxies have
L$_{IR}$/L$_B$ $> 10$. In general, the HII galaxies have higher L$_{IR}$/L$_B$
ratios than
the SBNGs.
Most of the SBNGs have a ratio coherent with a near constant star formation
rate
over the last 3 Gyr.}
  \label{Fig.7}
  \end{figure*}

Using L$_{IR}$ and L$_B$ as two tracers of SFRs on different time
scales, we now proceed to compare the histories of star formation of
the various types of star forming--galaxies. To quantify our
luminosity-luminosity diagram in terms of SFRs, we have adopted the
following model. We simplify the stellar formation process by
separating it in two independent functions (Matteuci 1989): $B(m,t)
= \psi(t) \phi(t)$, where $\psi(t)$ is the SFR and $\phi(t)$ is the
IMF. Parametrizing further the SFR as a an exponential function:
$\psi(t) \propto exp(-t/\tau)$, where $\tau$ is the characteristic
time scale of the star formation. Following this model, short time
scales correspond to small values of $\tau$, and a constant SFR
corresponds to $\tau \rightarrow \infty$. We use a Salpeter's IMF
with an upper mass limit of 100 M$_{\odot}$ and a lower mass limit of
0.1 M$_{\odot}$. Following this model, a mean constant SFR over the
last 3 Gyr period should yield a ratio L$_{IR}$/L$_B$ = 1 (Ghallager
\& Hunter 1987).

In Fig. 6, we test this model by comparing the  sample of active
spiral galaxies from Kennicutt (1983) with the sample of early-type
galaxies from Hogg et al. (1993).  Following Ghallager et al. (1984)
and Kennicutt (1983), late-type spirals have produced new stars at
roughly constant rates over the last 3 Gyr period (see also
Kennicutt et al. 1994). While rapid star-formation has lead to
early-type galaxies (Charlot \& Bruzual 1991), which have now very
low SFR compared to their past values. In Fig. 6, most of the
late-type spirals are within a factor 3 of the constant SFR line,
while most of the early-type galaxies are clearly below this line.
In this paper, we will consider that in a galaxy with L$_{IR}$/L$_B
> 3$ the SFR has probably increased by a substantial factor over
its historical mean value, while in a galaxy with L$_{IR}$/L$_B <
1/3$ the present SFR has significantly declined.

In Fig. 6, we have also
included the sample of Blue Compact Dwarfs (BCDs) and Irregular
galaxies (Irs) from Thronson \& Telesco (1986), and a sample of
starburst IRAS galaxies from Allen et al. (1991). Following the adopted model,
most of the IRAS starbursts have a present SFR much higher than
their mean past value. The BCDs and Irs also seem to have higher
SFR, although at a lower level than the IRAS starburst.

In Fig. 7, we compare the histories of star formation of the
different types of starburst galaxies.  The HII galaxies are
from the UM survey (Salzer et al. 1989) and the Calan-tololo survey (Pe\~na et
al. 1991).
The SBNGs are composed of the galaxies from the UM survey, the Balzano's sample
(1983)
and from the MBG survey (Coziol et al. 1993; Coziol et al. 1994). The last
sample
is composed of the HII nucleus galaxies (French 1980; Heckman 1980;
Stauffer 1983).
Comparison of Fig. 6 with Fig. 7 shows that the HII galaxies, the BCDs and the
Irs
occupy the same region of the diagram.
As we have stated earlier, they are of the same nature.
As compared to the SBNGs, these samples of galaxies seem to contain a
higher fraction of more active starburst galaxies.  In part, this
difference can be explained by the methods used to detect these
objects: surveys based on the presence of lines are slightly biased
toward higher active galaxies than surveys based on UV-bright excess.
In the case of dwarf galaxies, it is also easier to pick out mostly
bright galaxies, which turns out to be also the most active star
forming galaxies of the population (Ferguson 1993; Melisse \& Israel
1994).

\begin{figure*}
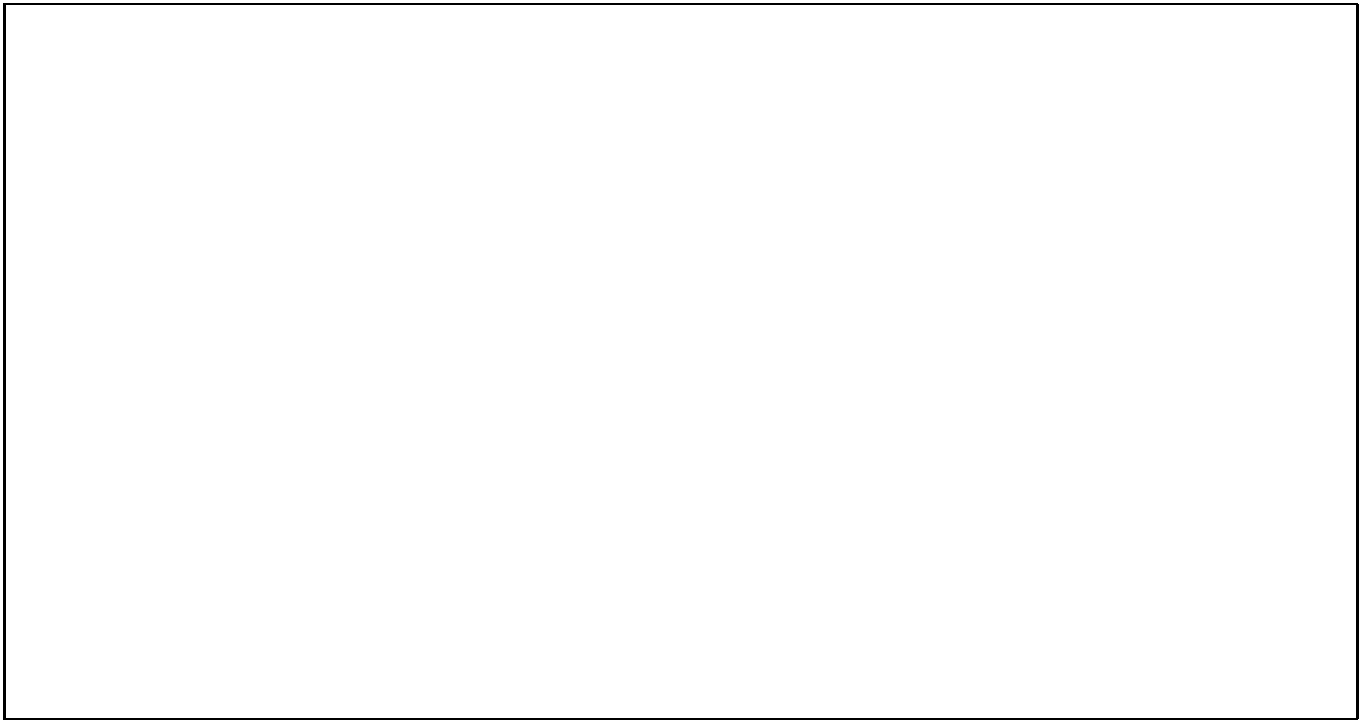

  \picplace{9.5cm}
  \caption{The luminosity-luminosity diagram for the AGNs. We descriminate two
types of LINERs: the
optical LINERs are located in spirals, with present low star formation rates,
while
the IRAS LINERs are in massive starbursts. The position of the ultra--luminous
infrared galaxies
(galaxies with L$_{IR}$/L$_B > 10$) as compared to
the Seyfert galaxies suggest that they
could be dust rich embedded AGNs associated to a strong starburst.}
  \label{Fig.8}
  \end{figure*}

In general, the HII galaxies trace a continuous sequence of luminosity with the
SBNGs. This sequence can be interpreted as a sequence in stellar mass.
The more massive galaxies (that is, galaxies with L$_B > 10^9$ L$_\odot$),
seem to trace a continuous sequence of SFRs. The SFR increases from
the HII nucleus galaxies to the SBNGs, and up to the IRAS starburst.
Perhaps the most
important feature of Fig. 7 is that most of the starburst galaxies
are found in the region where the ratio L$_{IR}$/L$_B$ is within a
factor 3 of the line of constant SFR over a 3 Gyr period.  There is
very few ultra-luminous infrared galaxies (defined here as galaxies with a
ratio L$_{IR}$/L$_B > 10$).

For comparison, we have included in Fig. 7,
some well known starburst from the literature. M82 is one
of the most studied starburst galaxy. It is commonly know as the
``typical starburst''. In Fig. 7, this galaxy looks more like an extreme case.
 II Zw 40 is another extreme case.  In our samples, this galaxy is
one of the rare examples of a very active dwarf. On the other hand,
NGC 7714 deserves its epithet of the archetype of the SBNGs
(Weedman et al. 1981). Mrk 710 is one example of the variety of
galaxies that are called starburst. This galaxy is directly on the
line indicating constant SFR over a Gyr period.

Arp 220, NGC 6240, NGC 3690 and NGC 1614 are all examples
of ultra-luminous infrared galaxies (Sanders et al. 1988;
Armus et al. 1989). In Fig. 7, the position of the IRAS starbursts is similar
to
those of the ultra-luminous infrared galaxies, which suggests a similar nature.
Actually, the origin of the activity in those galaxies is not very
clear. In fact, some of the ultra-luminous infrared galaxies shows
optical spectra similar to a Seyfert 2 or of a LINER. Therefore, if
an AGN is also present in those galaxies, and contribute to heat
the dust, the interpretation of the ratio L$_{IR}$/L$_B$ in terms
of different SFRs is not totally correct.   To examine this
question further, we compare in Fig. 8 the luminosities  of the
IRAS starburst galaxies with those of a sample of Seyfert 2
galaxies (Bonnatto \& Pastoriza 1992), and of two different samples
of LINERs: the original sample of LINERs detected in the optical by
Heckman (Heckman 1980), and a sample of IRAS LINERs (Allen et al.
1991).

The position of the optical LINERs in Fig. 8
suggests that they reside in galaxies with declining SFRs. On the contrary,
all the IRAS LINERs seem to be in a very strong starburst phase.
This dichotomy between the two samples of LINERs suggests that we
should be careful when we try to determine which phenomenon best
explain the optical spectrum of a LINER, in particular, between
photoionization of a nonstellar continuum or shock heating related
to an ongoing starburst (Ho et al. 1993). By comparison, the
Seyfert 2 galaxies have much lower ratios L$_{IR}$/L$_B$ than the
IRAS starburst (note that in this kind of diagram, the Seyfert 1
galaxies are indistinguishable from the Seyfert 2). In Fig. 8,
either the AGN do not contributes to heat the dust, and all the
Seyfert 2 galaxies are also in a starburst phase, or the AGN
contributes significantly and the Seyfert 2 reside in low-star
forming galaxies. The problem is probably more complicated as it
should also consider the amount of obscuring material collected in
the nuclear region (Maiolino et al. 1995).

\begin{table*} \caption{Massive stars characteristics}
\begin{tabular}{cccccc} \hline\noalign{\smallskip}
spectral&Mass&L$_{bol}$&N$_L$&L$_{H\alpha}$&L$_{H\alpha}$/L$_{bol}$\\
type&(M$_{\odot}$)&(L$_{\odot}$)&(s$^{-1}$)&(L$_{\odot}$)&\\
\noalign{\smallskip}\hline\noalign{\smallskip}
O5   & 49 & $6.8 \times 10^5$ & $4.2 \times 10^{49}$ & $1.5 \times
10^4$ & $2.2 \times 10^{-2}$ \\ O9   & 22 & $4.6 \times 10^4$ &
$1.2 \times 10^{48}$ & $4.3 \times 10^2$ & $9.3 \times 10^{-3}$ \\
B0   & 19 & $2.5 \times 10^4$ & $2.3 \times 10^{47}$ & $8.3 \times
10^1$ & $3.3 \times 10^{-3}$ \\ B0.5 & 15 & $1.1 \times 10^4$ &
$1.7 \times 10^{46}$ & $6.1 \times 10^0$ & $5.5 \times 10^{-4}$ \\
B1   & 12 & $5.2 \times 10^3$ & $1.9 \times 10^{45}$ & $6.8 \times
10^{-1}$ & $1.3 \times 10^{-4}$ \\ \noalign{\smallskip}\hline
\end{tabular} \end{table*}

By comparing the IRAS starbursts with AGNs, we can conclude that
the IRAS LINERs and the IRAS starburst galaxies could have a very
special nature. Those galaxies could be dust rich embedded AGNs,
associated with a strong starburst. Consequently, the HII galaxies
and the SBNGs are probably more representative of the starburst
phenomenon. Therefore, the  most important characteristic of a
starburst galaxy should be
 a mean constant SFR over a few Gyr period.

\section{The difference between HII galaxies and SBNGs}

Following our analysis, the principal difference between the HII galaxies and
SBNGs
resides in their mean ratios L$_{H\alpha}$/L$_B$ and L$_{H\alpha}$/L$_{IR}$.
This difference of ratios explain why, in Fig. 3 and in Fig. 4,
the SBNGs do not occupy the same region of the diagram as the HII
galaxies and the disk HII regions. In SBNGs both ratios are near
1000 while in the HII galaxies and in the late-type spirals, the
same ratios are near 100. We have also earlier that this
phenomenon cannot be explained by a higher dust extinction in the SBNGs.
In this section we explore further this problem by comparing the energy
budget of the different starburst galaxies.

\begin{figure}
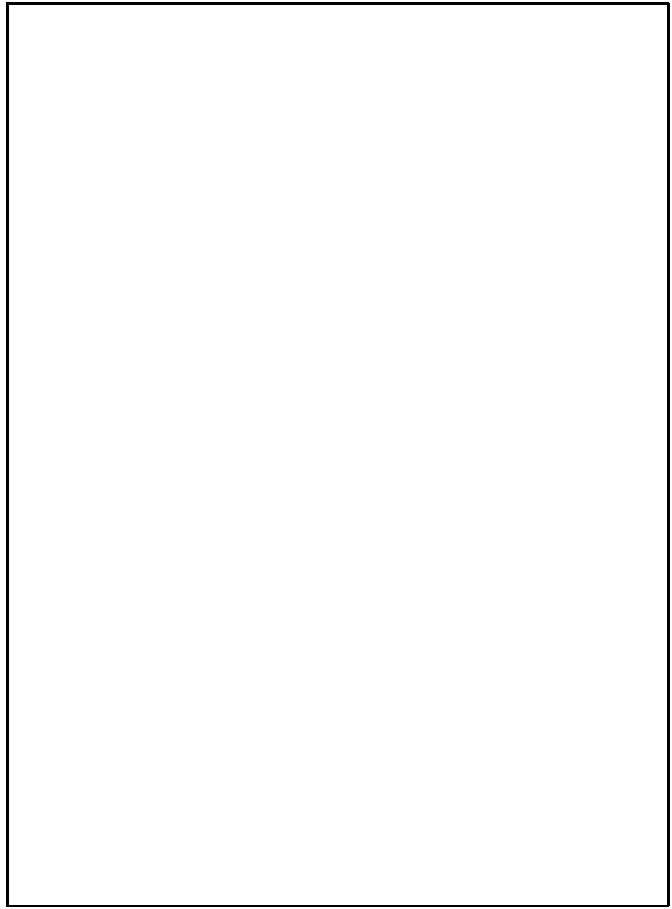

  \picplace{12cm}
  \caption{The energy budget of starburst galaxies. The continuous lines give
the ratios
between the ionizing luminosity and the bolometric luminosity for stars of
different spectral types.
The vector R corresponds to the correction between the bolometric luminosity
and the far--infrared luminosity (f), and the correction for an optical
extinction Av $=$ 1 magnitude.}
  \label{Fig.9}
  \end{figure}

The energy budget of ionized regions can be checked by comparing
the ratio of L$_{H\alpha}$ with L$_{IR}$.
The idea behind this method is that the energy produced by a
star to ionized the gas is also sufficient to heat the dust which
emits in the far--infrared band (Devereux \& Young 1990;
Calzetti et al. 1995). Therefore, the different
ratios L$_{H\alpha}$/L$_{IR}$ should be comparable to the ratio of the
H$\alpha$
luminosity to bolometric luminosity (L$_{bol}$) of ionizing stars of different
spectral types
(Devereux \& Young 1990).
Those ratios can be found in column 6 of table 3, together with the parameters
used
to evaluated them. In column 1, we have the spectral type of the
stars. In column 2, the masses of the stars were calculated using
the relation (Allen 1973): Log(L/L$_\odot$)$ = 3.45$
Log(M/M$_\odot$). Columns 3 and 4 gives respectively the bolometric
luminosities and the number of ionizing photons, as calculated by
Panagia (1973). The H$\alpha$ luminosity in column 5 was calculated
using the relation: N$_L = 2.2$ L$_{H\alpha}$/ $3.03 \times
10^{-12}$ s$^{-1}$.

In Fig. 9, overplotted on the L$_{H\alpha}$, L$_{IR}$ diagram, we
have reported the ratios from column 6 of table 3. Note that the
different ratios produce lines, because they correspond in fact to
cluster of stars. In Fig. 9, the f vector  is a correcting factor
between the far--infrared luminosity and the bolometric luminosity
(Devereux \& Young 1990). In this figure, we
 give also the vector corresponding to a correction of 1 magnitude dust
extinction. A 1 magnitude dust extinction is the mean
value usually evaluated in starburst galaxies by the Balmer decrement
(Balzano 1983; Pe\~na et al 1991; Masegosa et al.
1994; Calzetti et al. 1995).  As a byproduct, this kind of diagram
could also indicates the equivalent spectral type of the ionizing
stars. Taken at face value, Fig. 9 suggests that the HII
regions of SBNGs are dominated by B type stars.

\begin{figure}
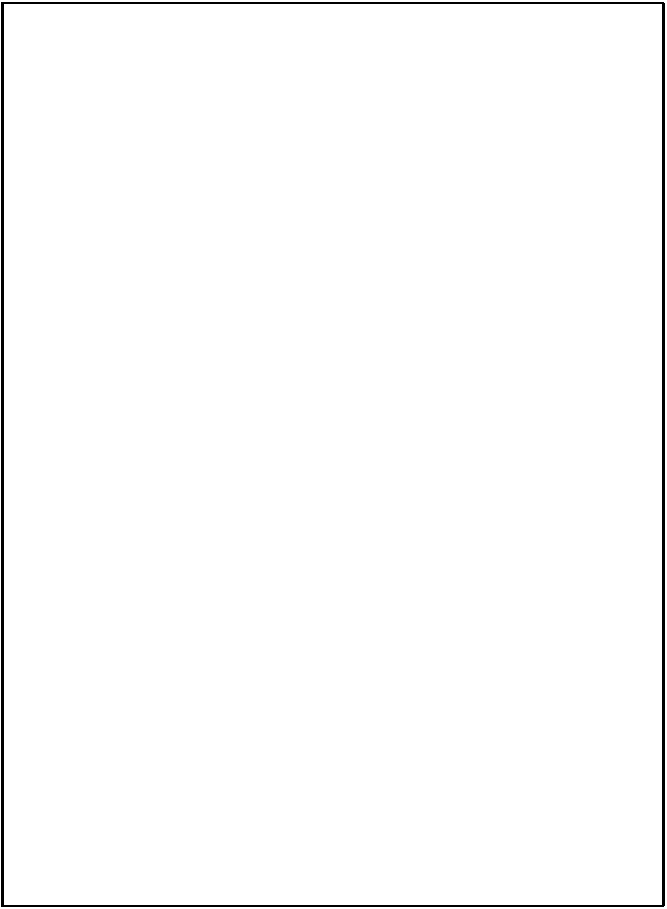

  \picplace{12cm}
  \caption{The energy budget of the SBNGs considering
the maximum number of corrections as suggested by CBKH (see details in the
text). The starburst galaxies
from CBJKH have been grouped
following their level of extinction, as determined by the UV
spectral index $\beta$. The position of a galaxy in this diagram do not depends
on its level of extinction.}
  \label{Fig.10}
  \end{figure}

In the literature, we can find similar reports of this phenomenon.
For example, in their study of the UM galaxies,  Salzer et al.
(1989) have found that if we use the B
luminosity to determine the number of O7 star present in the
starburst galaxies, the ratio of massive stars decreases in the SBNGs as
compared
to the HII galaxies (see their figure 8). In
their study of the Markarian SBNGs, Deutsch \& Willner (1986) have
also observed a better correlation between L$_B$ amd L$_{IR}$,
which they explained by a greater contribution of B type stars to
heat the dust.  Finally, by modeling the Bracket lines of a sample of
starburst galaxies, Doyon et al. (1992) have found values suggesting
a higher number of intermediate-mass stars, which they interpreted as
a sign of a truncation of the IMF of starburst towards
intermediate-mass stars (they proposed an upper-mass limit of, at
most,  30 M$_\odot$, which is closer to an O9 star than an O7 star,
see Table 3).

In their study of starburst galaxies, CBKSH have also determined the
energy budget of their sample of starburst galaxies. One of their conclusion is
that the
stellar populations responsible for the observed UV and optical
spectrum is sufficient to explain the far--infrared fluxes in the IRAS bands.
However, they
have determined that only 70\% of the IRAS flux could be attributed
to warm dust. To fit their model in the far--infrared, they had also to suppose
that 30\% of the warm
dust was due to massive stars embedded in dusty regions,
which were not manifest through UV or optical emission. In Fig. 10,
we have applied all their corrections to our sample of SBNGs, as well as
to the sample of starbursts of CBKSH. All the galaxies were also
corrected for internal extinction, supposing a 1 magnitude extinction
for the SBNGs. In Fig. 10, even with the corrections, we cannot
change the observed ratios L$_{H\alpha}$/L$_{IR}$ to make them
compatible with stars more massive than spectral type O5. The sample
of galaxies of CBKSH seems to work better, because it is slightly
biased towards more active galaxies (as judge from the mean far--infrared
luminosity Log$<$L$_{IR}>$
= 10.2). In Fig. 10, we have grouped the starburst of CBKSH
following their level of extinction, as determined by the UV
spectral index $\beta$. Increasing amount of dust obscuration
corresponds to increasing values of $\beta$. As can be seen, the
position of a galaxy in Fig. 10 does not depends on the amount of
extinction.

\vskip -0.5cm

\vskip 0.5cm

To reproduce ratios L$_{H\alpha}$/L$_{IR}$ compatible with those of
massive stars of spectral type O5 and earlier, we have to
suppose a mean optical extinction of 3 magnitudes. This however is
inconsistent with the mean extinction value, as estimated by the
Balmer decrements or the UV spectral index (Balzano 1983; CBKSH).
A mean correction of 3 magnitudes would also implies an even greater
correction of the B luminosity (by about a factor 3), because the B
band is more affected by extinction than the V band. The corrected bolometric
luminosities of the SBNGs would then be much more higher than
normal HII regions, and comparable to those of nearby AGNs. But,
such luminous objects surely would gives other signs of their high
activity. In the case of SBNGs, we do not see those signs. In fact,
the only possible case of extreme dust rich starburst galaxies in
our samples are the IRAS starburst or the ultra-luminous infrared
galaxies.
But, all these galaxies produce far--infrared emission much higher than the
SBNGs.

We can also suppose that in SBNGs less than 70\% of the far--infrared
emission comes from the hot dust. Sauvage \& Thuan (1992) have
determined that between 50\% to 80\% of the far--infrared emission
is due to cool dust, depending on the morphology of the galaxy.
Considering that the mean temperature of the dust in SBNGs is much
higher than in the late-type galaxies (see Fig. 2), the value of
30\%, as determined in the model of CBKSH, is probably the lowest
limit possible for starburst galaxies. We could also suppose a
higher ratio of non-ionizing stars. However, to make the SBNGs
similar to the HII galaxies, we would have to suppose that as much
as 90\% of the stars are invisible in the optical and in the UV.

The main difference between the HII galaxies and
the SBNGs could be a difference of stellar population. Supporting
this interpretation, we show in Fig. 11, the distribution of the H$_\alpha$
equivalent width (W$_{\lambda}$(H$\alpha$)) of the different types
of starburst galaxies. The equivalent width of the H$_\alpha$ emission
line is useful to estimate a relative population of massive to
less-massive stars, because H$_\alpha$ traces the highest mass
stars and the continuum at the wavelength of H$_\alpha$ traces
lower mass stars. Despite strong dependences on the metallicity,
the upper-mass limit or the slope of the IMF, the equivalent width
is basically determined by the evolution of the HII regions (Copetti
et al. 1986): it decreases with the age of the HII region. From Fig.
11, it is clear that the stellar population of the SBNGs is
much different than from those of the HII galaxies. The dashed line in
Fig. 11 corresponds to the median value for normal disk HII regions
(Kennicutt et al. 1989). The equivalent width of the metal poor HII
galaxies are comparable to those of normal disk
HII regions. The equivalent width is 10 times
lower in SBNGs, and almost a 100 times lower in the less active HII
nucleus galaxies.

\begin{figure}
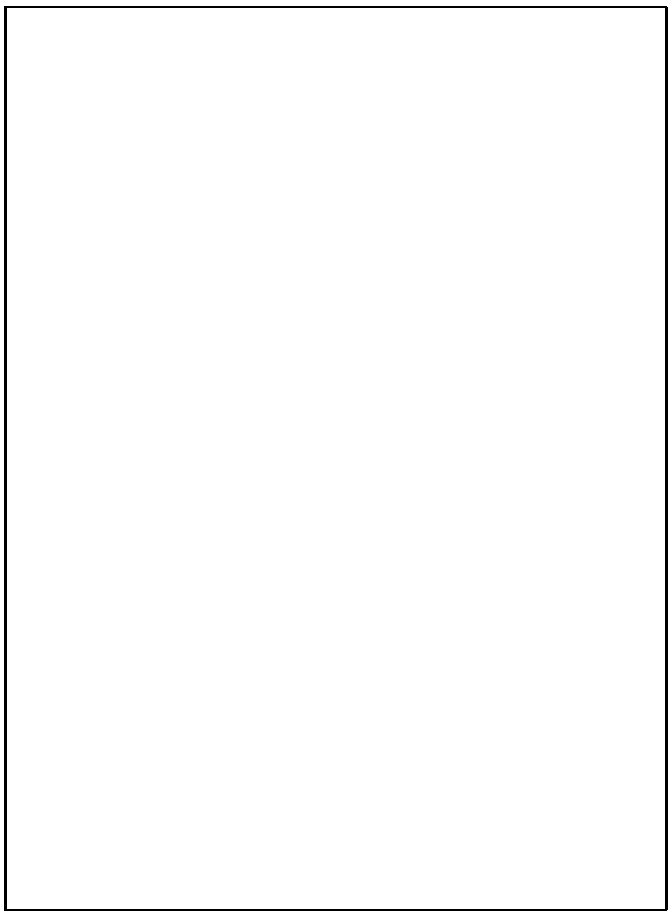

  \picplace{12cm}
  \caption{Comparison of the dispersion of the equivalent width of
H$\alpha$ (W$_{\lambda}$(H$\alpha$)) for the two types of starburst galaxies.
The dashed line corresponds to the median value for normal disk HII regions.
The equivalent width is 10 times
lower for the SBNGs than for the HII galaxies, which suggest different burst
populations.}
  \label{Fig.11}
  \end{figure}
\vskip -0.5cm

\section{The regulating role of massive stars in the evolution of
galaxies}

We see two possibilities to explain the difference of stellar
populations between SBNGs and HII galaxies.  The first one is to
suppose that in SBNGs the IMF is truncated towards intermediate-mass stars
(Doyon et al. 1992). The second possibility is that the HII regions
of SBNGs are more evolved on the average than in normal HII regions.
Without further observations, we cannot discriminate what is the
exact solution. The main reason is that we still do not know the
details of how stars are formed, and in particular, if their
formation implies a universal IMF. However, we believe that the
actual observations, based on studies of molecular clouds and of
star formation in external galaxies, are in better agreement with
an universal IMF (Kennicutt et al. 1994).  Furthermore, in luminous
starburst it is more often proposed that the IMF is truncated toward massive
stars (Rieke et al. 1980;
Terlevich \& Melnick 1985; Rieke 1991).

The alternative to a variation of IMF is to consider that
the HII regions of SBNGs are more evolved than in HII galaxies. In
SBNGs, this would implies burst stellar population older than $10^7$
yr. This scenario is however incompatible with a near constant SFRs on a time
scale of 3 Gyrs. It is important to note that the method that we have used to
determined the
history of star formation of our galaxies is not sensitive to time
scale shorter than 1 Gyr (Gallagher \& Hunter 1987). Therefore, the
SFR of the starburst galaxies could have changed on shorter time
scales. In fact, it is possible to reproduce a mean constant star
formation rate on a few Gyr scale by a sequence of short bursts
(Coziol \& Demers 1995). To be able to distinguish some evolution
of the star forming regions, we would need bursts of star formation
with $\tau$ of the order of $10^7$ or $10^8$ (Charlot \& Bruzual
1991). The large population of intermediate-mass stars observed in
SBNGs would therefore suggests that the SBNGs have experienced a
succession of many short duration bursts.

Our observation are qualitatively consistent with the stochastic
self propagation star formation theory (SSPSF), as proposed by
Gerola et al. (1980). This theory was originally elaborated to
explain the burst of star formation in HII galaxies. As a main
feature, it predicts that the star formation in HII galaxies are
happening in a sequence of bursts separated by long periods of
time. The active agents are the massive stars: via supernovae
explosion, star formation is propagated or stopped in the near
neighborhood of an active region, depending on the matter
available. The probability of propagation of the star formation
increases with the mass of the galaxy. In small galaxies the
frequency of the bursts will be low, while in more massive galaxies
it will approach near constant star formation rates.  Furthermore, the
frequency of bursts could determined the level of chemical evolution
of a starburst galaxy. Therefore, SBNGs galaxies are more
evolved than HII galaxies because they have experienced more bursts
of star formation.

The SSPSF theory is also consistent with the luminosity biases
observed between our different samples. In low-mass galaxies,
the variation of luminosity between two bursts is very large, and
could reach as much a 7 magnitudes (Gerola et al. 1980).  Therefore,
it is more easy to detect an active HII galaxy than its
non-active counterpart. On the contrary, the luminosity of massive
galaxies with multiple bursts will reach a certain equilibrium, and
the majority of SBNGs will be found with a luminosity value as
predicted by near constant SFRs. Consequently, massive starbursts,
with unusually high SFRs, are better detected in the
far--infrared than in the blue.

\section{Summary and conclusion}

We have verified that the different luminosities of various star
forming galaxies can be used to trace SFRs over different time
scales. Consistent with this idea, the H$\alpha$
emission is better correlated to the far--infrared emission than to
the B emission. Furthermore, the amount of far--infrared emission
per unit of stellar mass increases linearly with the amount of
ionized gas per unit of stellar mass. Therefore, both L$_{H\alpha}$
and L$_{IR}$ are tracers of recent star formation rates on time scales
of $10^7$ or $10^8$ yr, whereas L$_B$ is related to
intermediate-mass stellar population, and traces SFRs on a time scale of a
few Gyr.

Using L$_{IR}$ and L$_B$, and a simple model of star formation, we
have compared the star formation histories of different samples of
star forming galaxies. We have shown that our model is consistent
within a factor 3 with the history of star formation of normal field
galaxies. Applying this model to the two types of starburst galaxies, we have
found that
most of them have a ratio L$_{IR}$/L$_B$ compatible with a mean SFR near
constant over a few
Gyr period.

The HII galaxies have comparable ratios than the SBNGs, and
trace a continuous sequence in mass with the SBNGs. The HII nucleus galaxies
represent the
low-luminosity end of the SBNGs.
There are very few ultra-luminous infrared starburst galaxies.  The
nature of these galaxies is not well determined. In particular, the
IRAS starburst and the IRAS LINERs could be dust--rich embedded AGNs
with a very strong starburst.

Comparing the different linear relations for the HII galaxies and
the SBNGs, we have found some interesting differences. The amount
of ionized gas per unit of stellar mass is lower in the
SBNGs than in the HII galaxies. Furthermore, the amount of far--infrared
emission per unit of ionized gas is higher in the SBNGs than in the
HII galaxies. We have verified that these  effects cannot be
explained by supposing higher dust extinction in the SBNGs.
To examined further these differences, we have
checked the energy budget of the different starbursts.
Our analysis suggests that the burst populations of the SBNGs are
dominated by B type stars. We see two possibilities to explain
the difference of stellar populations between the SBNGs and the HII
galaxies: either the IMF in SBNGs is truncated towards intermediate-mass
stars, or the HII regions of SBNGs are more evolved than in the HII
galaxies.

We argue that our observations are compatible with the idea of
multiple bursts of star formation in the SBNGs. This would explain
the great number of intermediate-mass stars in SBNGs. Our analysis
is qualitatively consistent with the SSPSF theory, which predicts that
the frequency of bursts increases with the mass of the galaxy. In more massive
galaxies, this
frequency approaches a near constant star formation rates. Following this
theory,
the frequency of bursts could also determined the level of chemical
evolution of a starburst galaxy, which could
explain why the SBNGs are more evolved than the HII galaxies.
However, it still remains to understand the
details of this mechanism, and to better determine
the importance of the starburst phase in the evolution of these galaxies.

The possibility that the
bursts are regulated by the evolution of massive stars and could be
prolongated over longer periods of time could explain the many cases
of relatively isolated starburst galaxies.
Another possibility related to the prolongated starburst phase, is perhaps the
chance to identify the initial event that gave birth to the nearby
population of isolated starburst galaxies. For example, considering a
redshift distribution z $< 0.03$  for the SBNGs, a crude
calculation show that the time elapsed between this event and the
interacting blue galaxies at z = 0.2 is only of the order of a few
Gyrs. This is comparable to the prolongated time scale of the starburst
galaxies. Therefore, the signs of past interacting events are not
necessarily extinct, and the population of distant blue galaxies
has not necessarily completely disappeared today.

\begin{acknowledgements}

I wish to thank Vicky Meadows for providing unpublished data on the
IRAS galaxies, and Clarissa Barth for reading and commenting this
work. I wish also to thank the referee, James Lequeux, for the
comments and suggestions that greatly helped to improve this paper.
This research has made use of the NASA/IPAC Extragalactic Database
(NED) which is operated by the Jet Propulsion Laboratory,
California Institute of Technology, under contract with the
National Aeronautics and Space Administration. The financial
support of the brazilian FAPESP ({\em Funda\c{c}\~ao de Amparo \`a
Pesquisea do Estado de S\~ao Paulo}), under contracts 94/3005--0 is
gratefully acknowledged.

\end{acknowledgements}

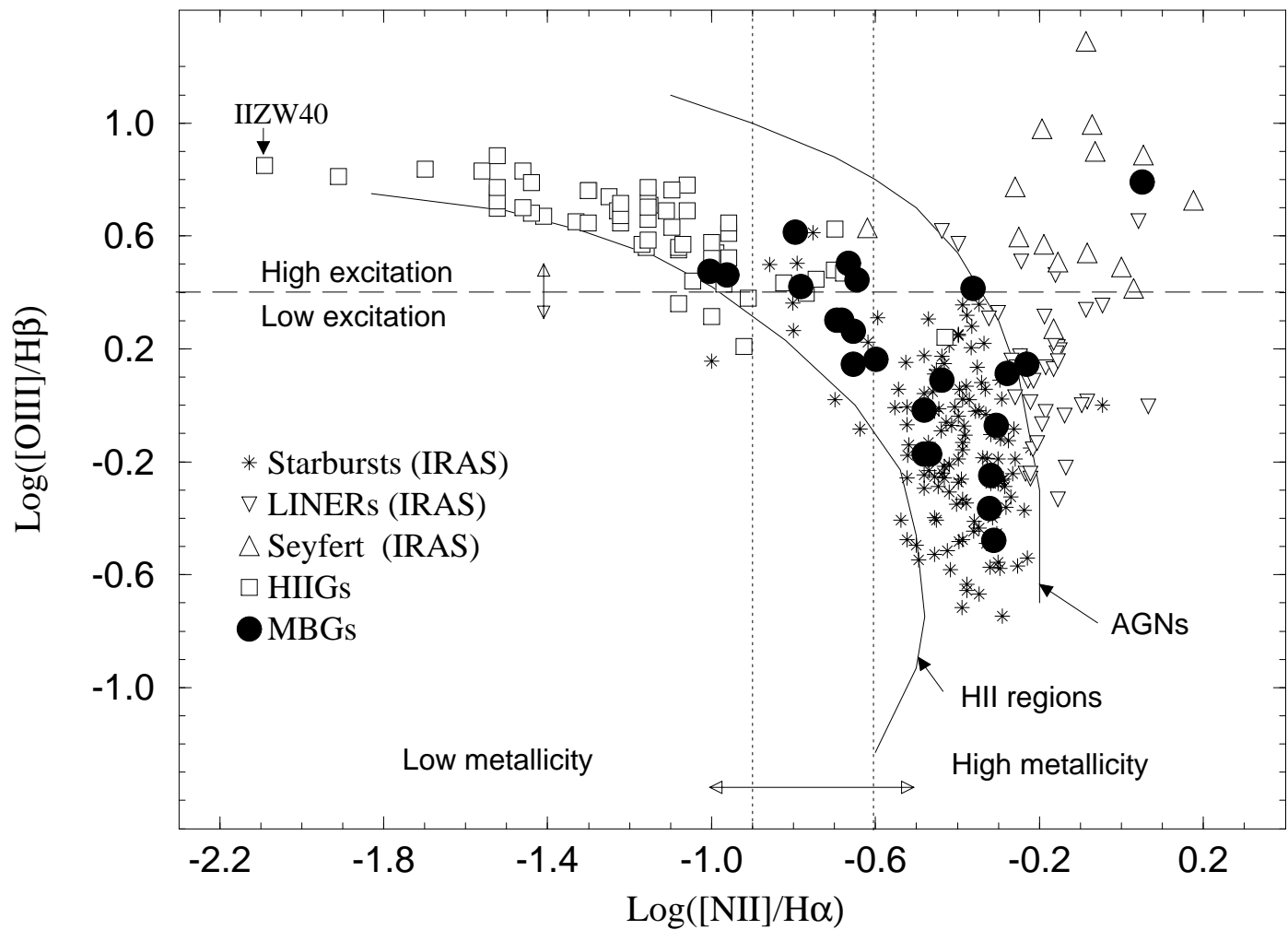
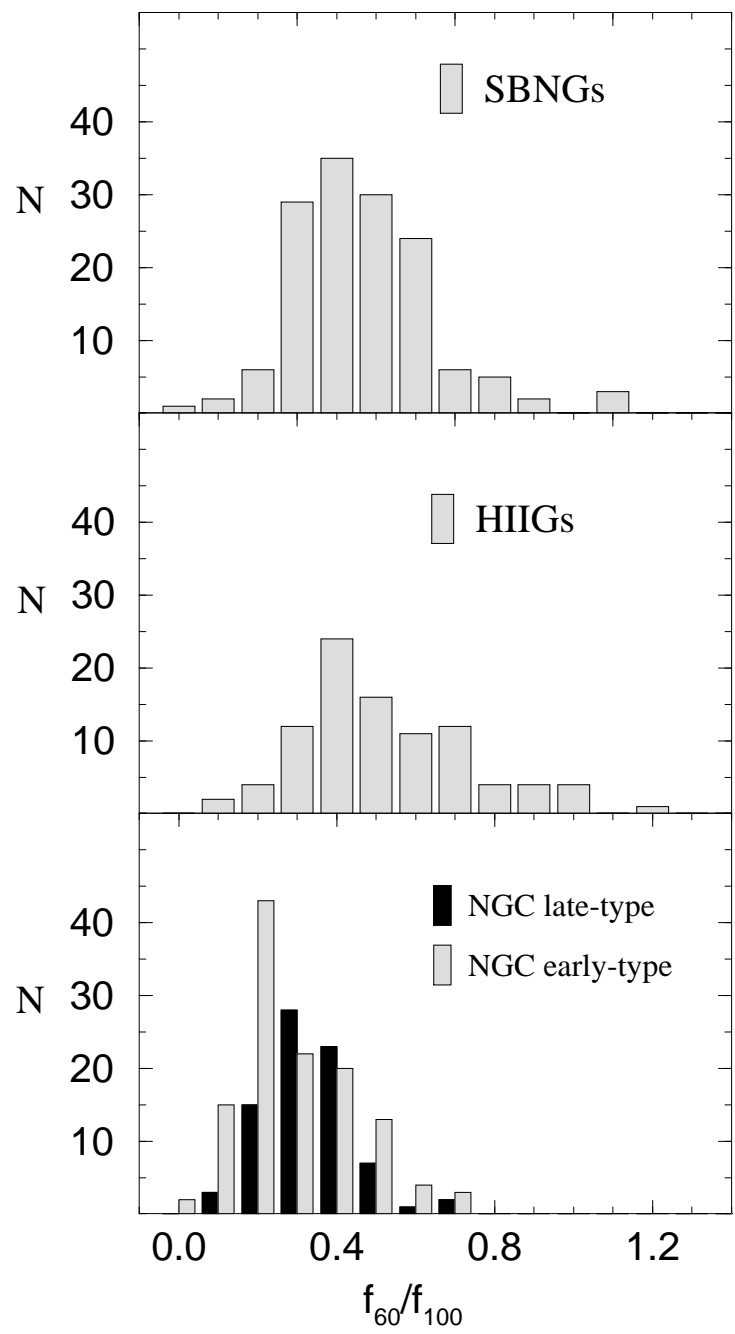
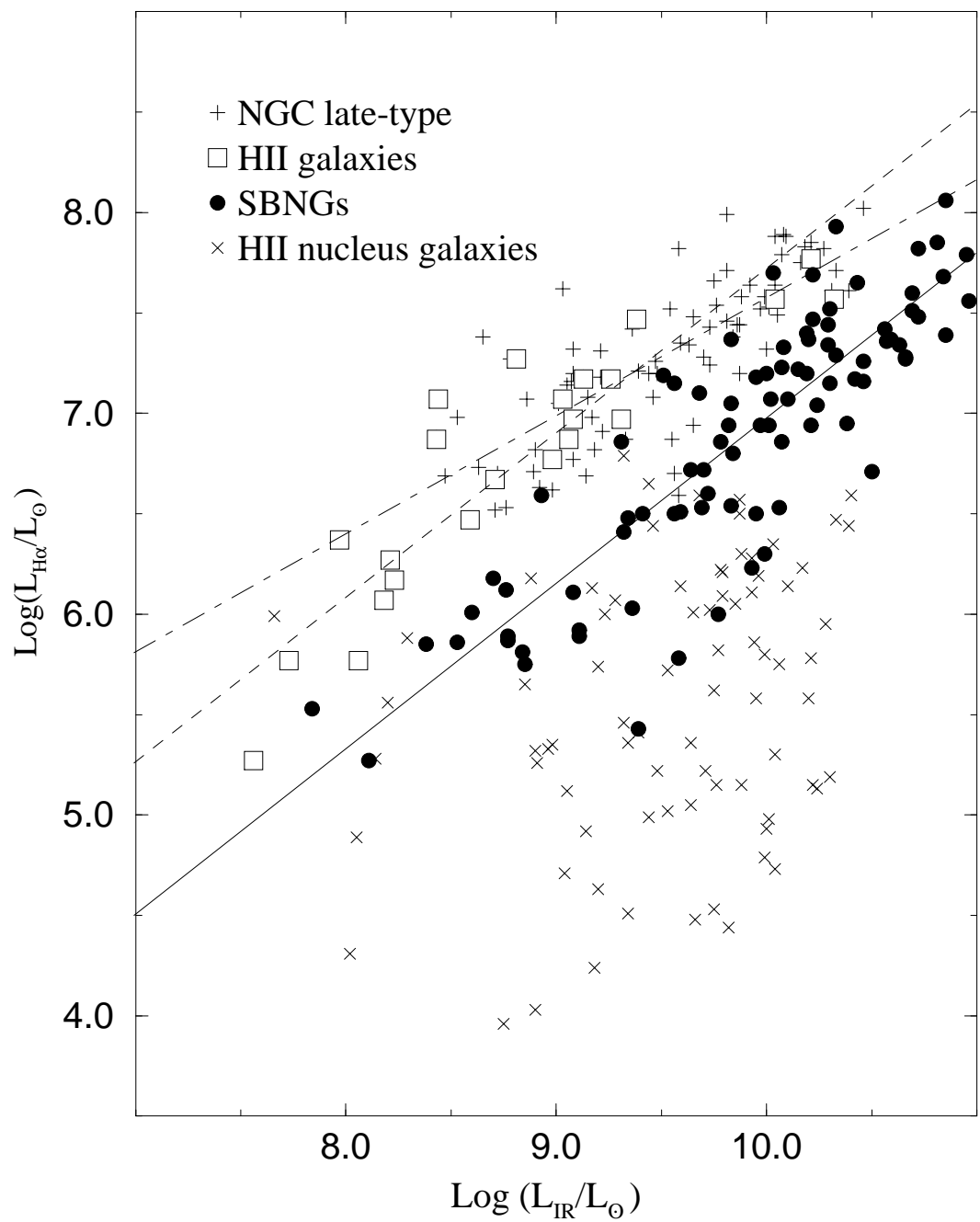
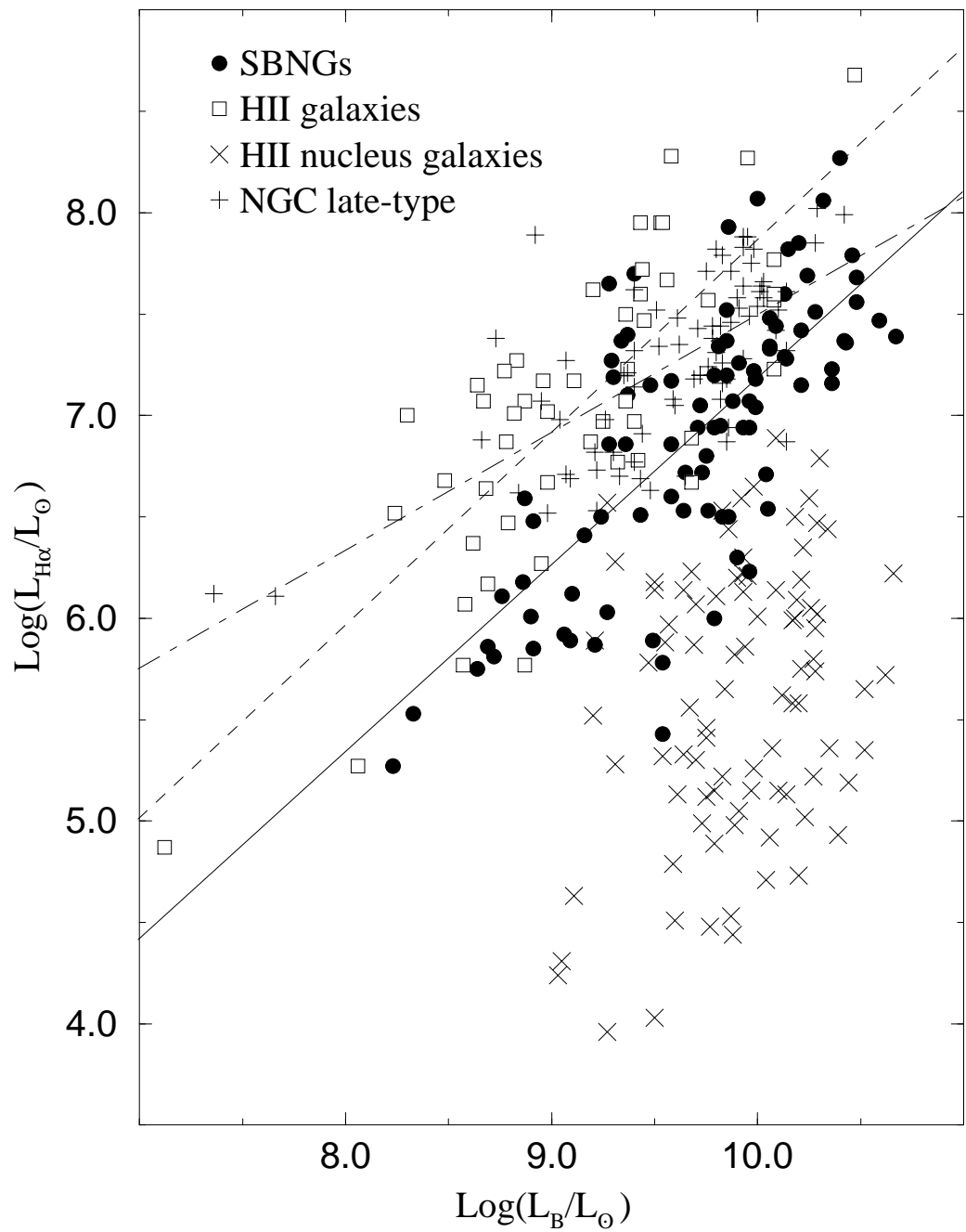
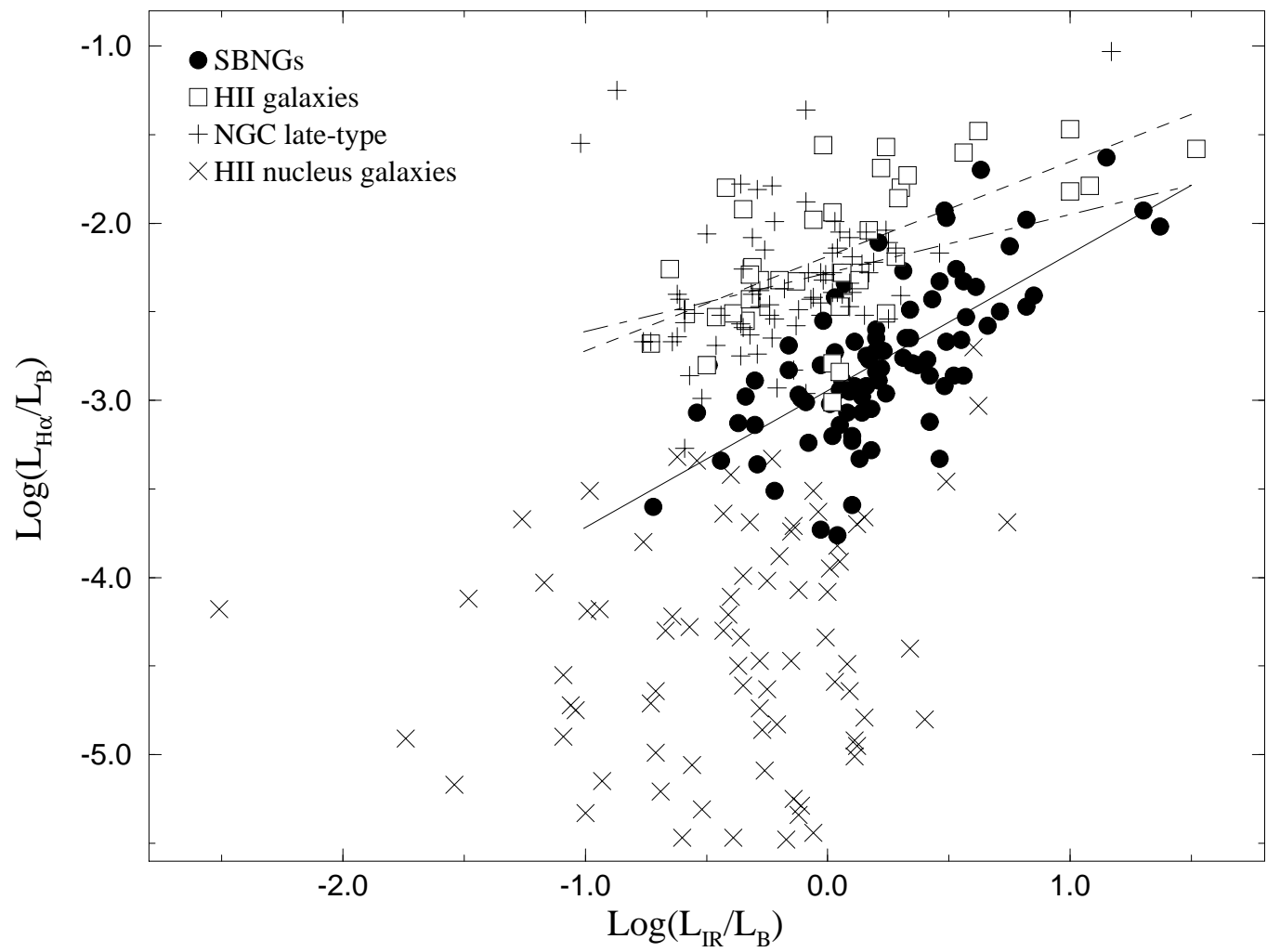
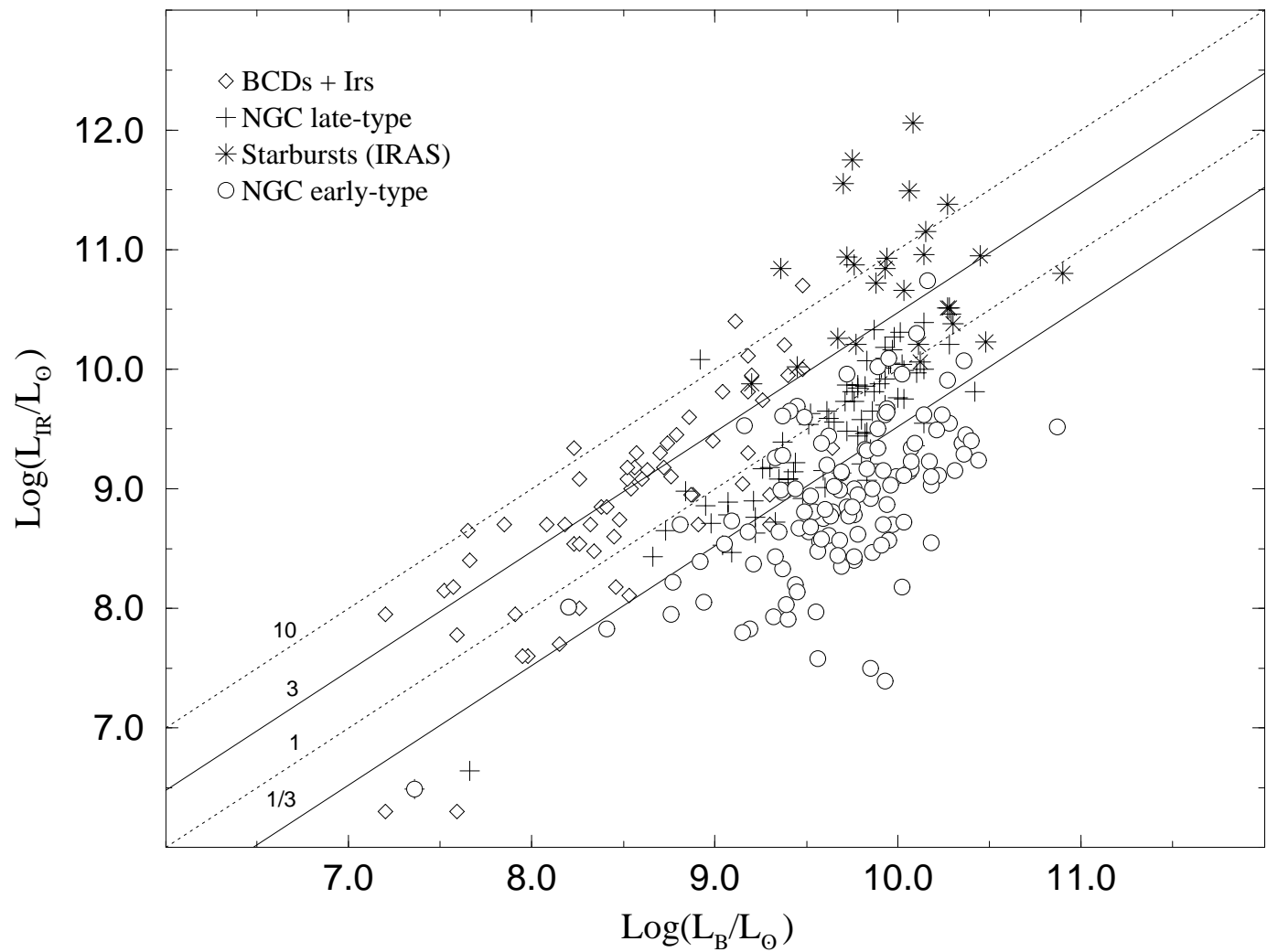
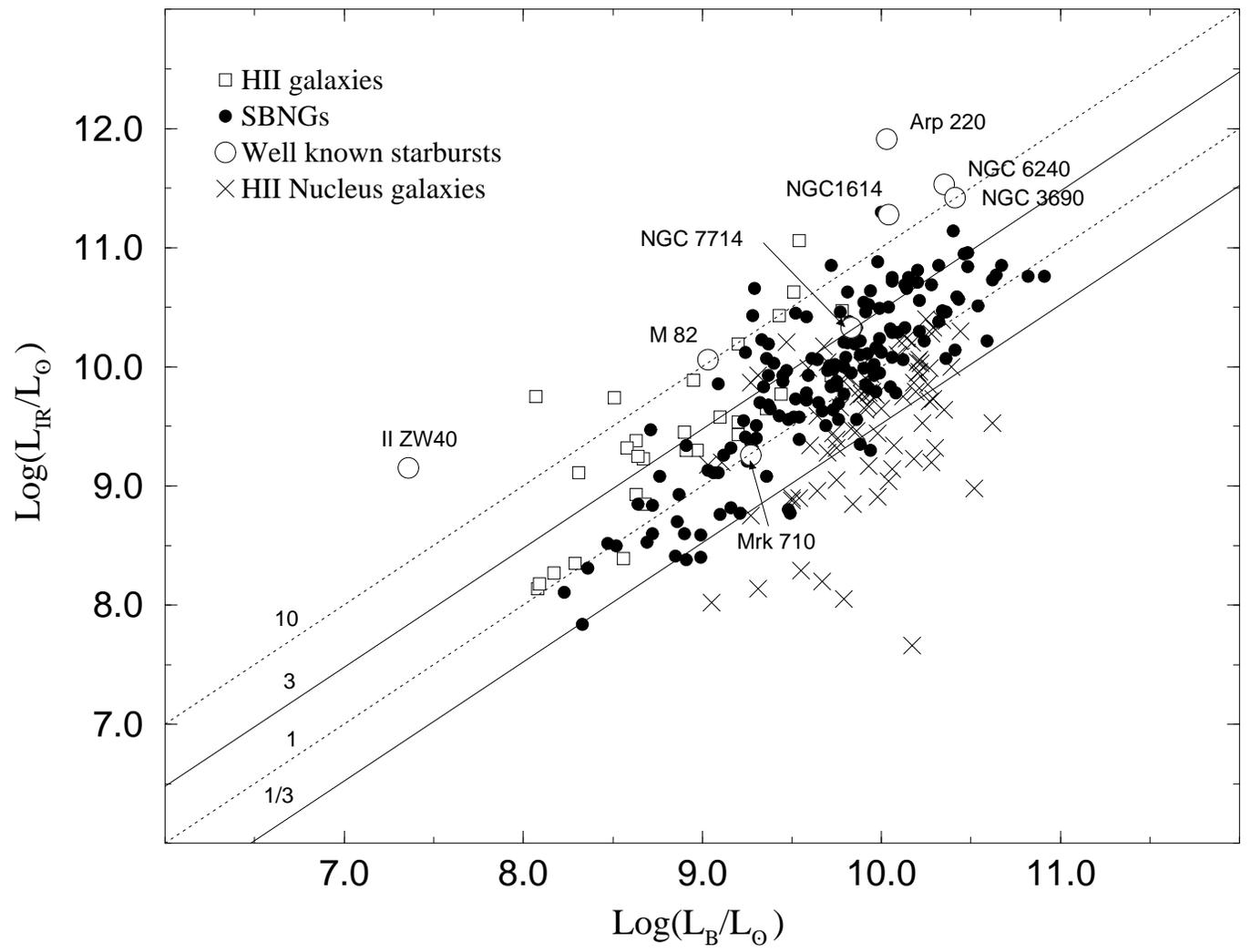
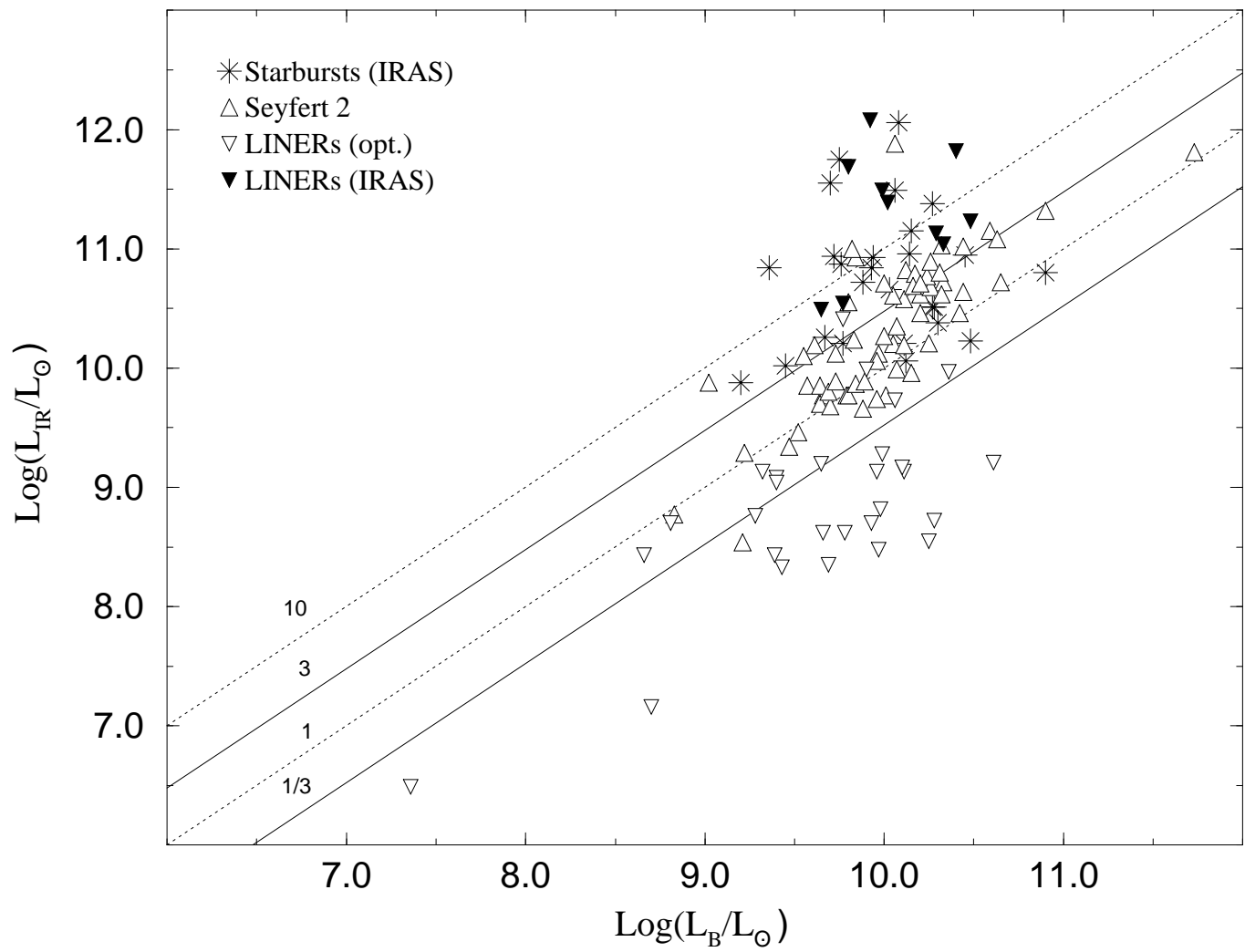
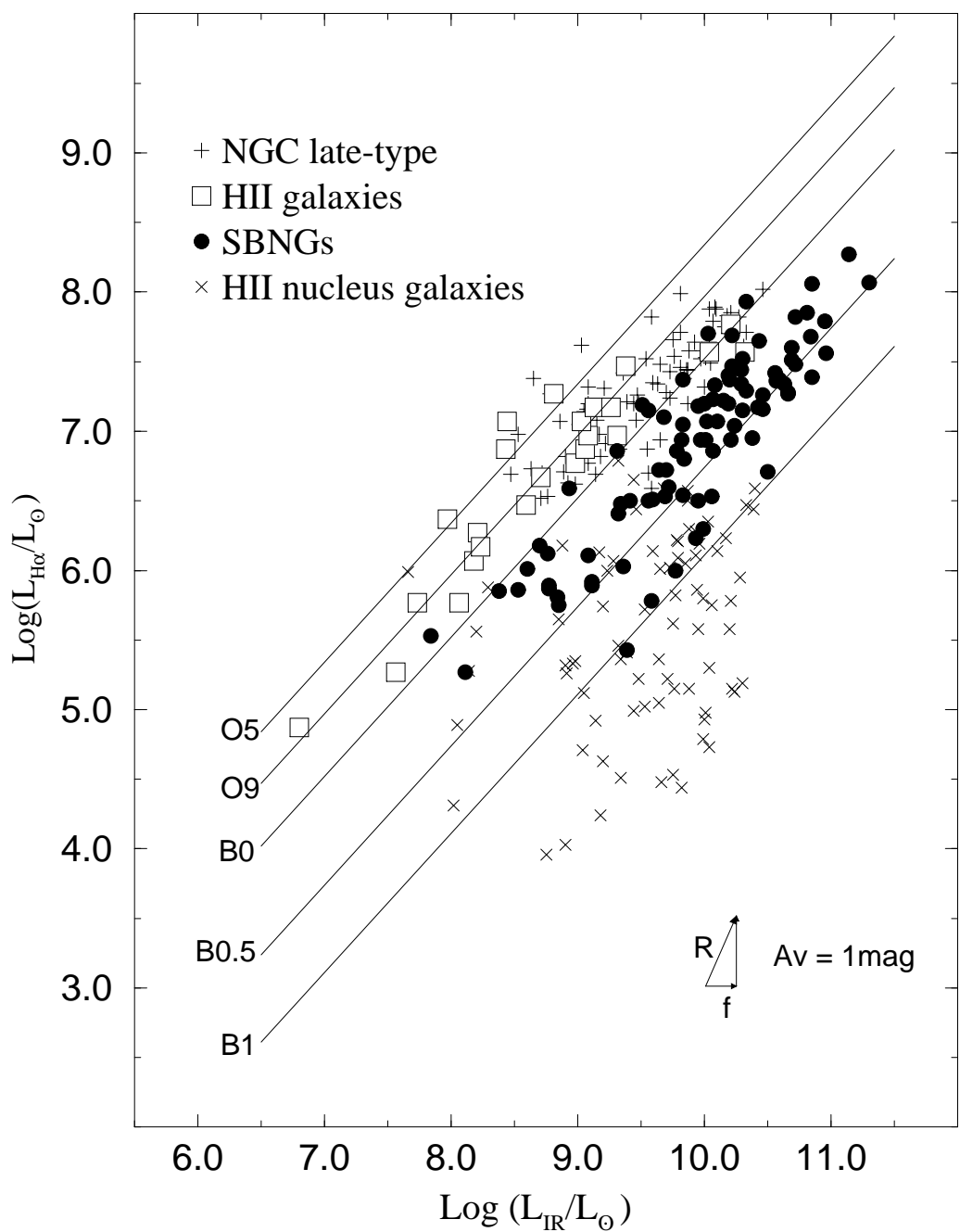
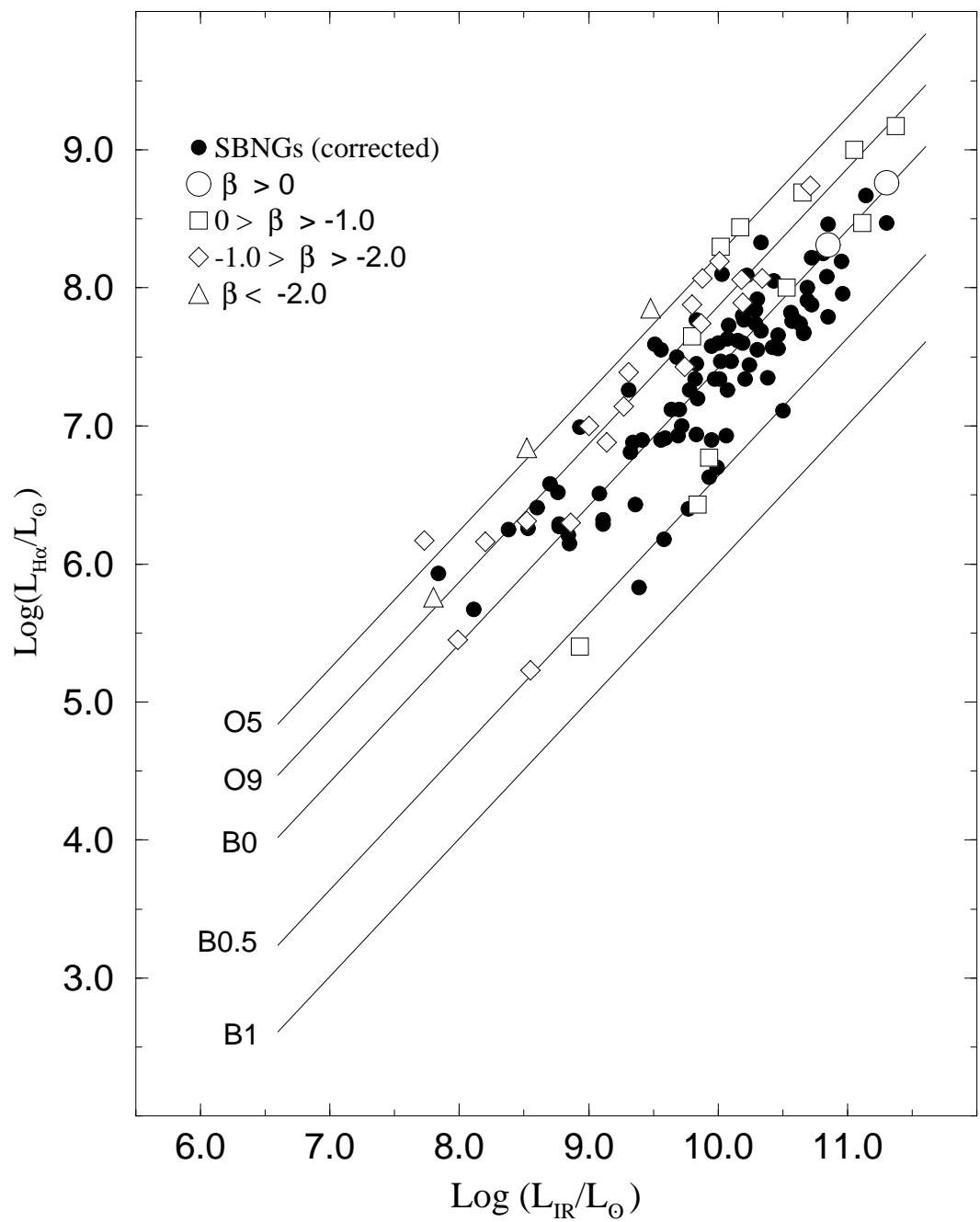
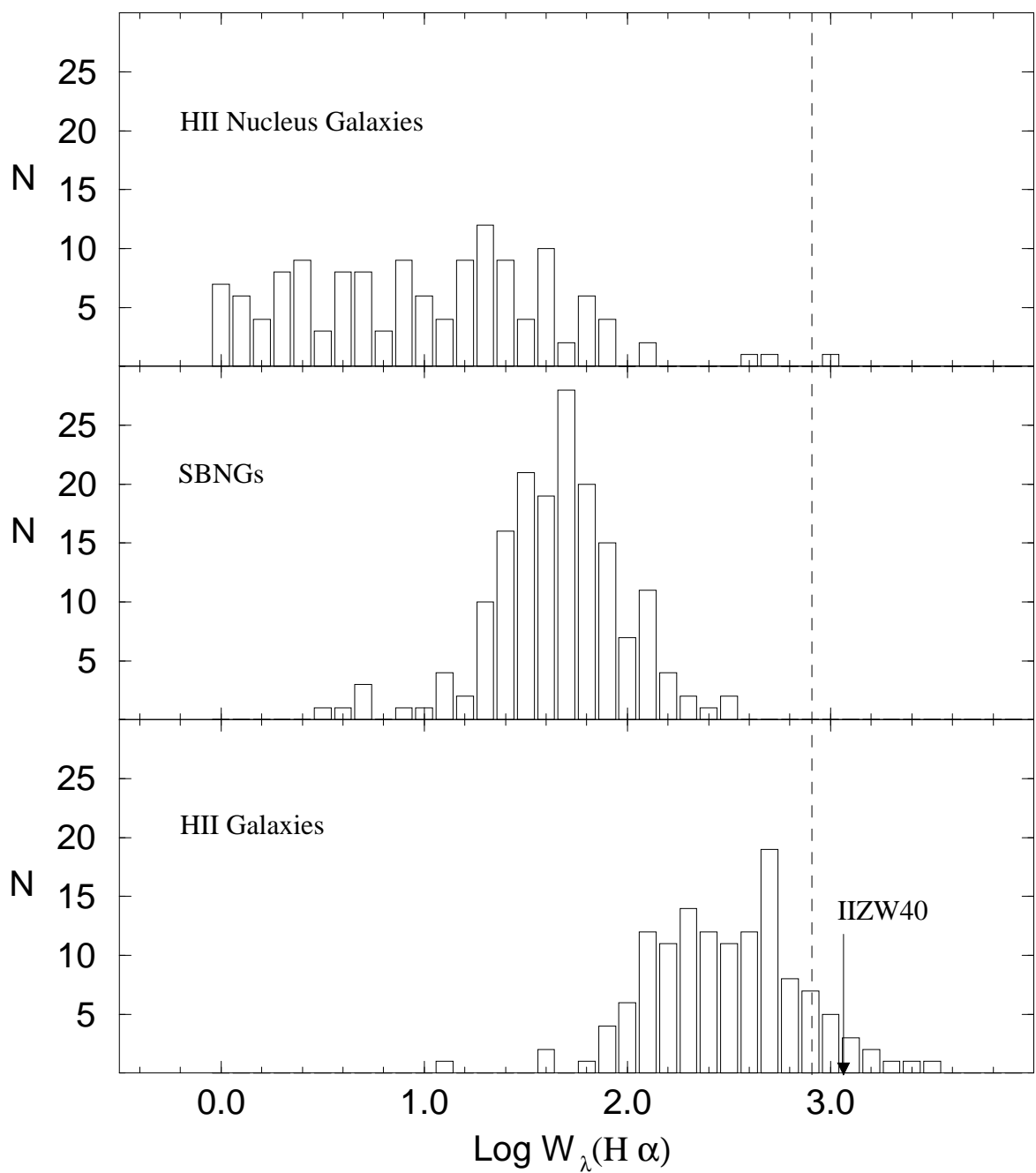
\end{document}